\pdfoutput=1

\documentclass[11pt,a4paper]{article}
\usepackage{jheppub}
\usepackage{slashed}
\usepackage{amsmath}
\usepackage{amssymb}
\usepackage{epsfig}

\newcommand{\labell} [1] {\label{#1}}

\def\({\left(} 
\def\){\right)}
\def\[{\left[} 
\def\]{\right]}
\newcommand{\non}{\nonumber \\}

\newcommand{\ie}{{\it i.e.,}\ }
\newcommand{\eg}{{\it e.g.,}\ }

\newcommand{\cN}{{\mathcal N}}

\newcommand{\A}{\mathcal{A}}

\newcommand{\cR}{{\mathcal R}}

\newcommand{\be}{\begin{equation}}
\newcommand{\ee}{\end{equation}}
\newcommand{\bea}{\begin{eqnarray}}
\newcommand{\eea}{\end{eqnarray}}

\newcommand{\mt}[1]{\textrm{\tiny #1}}

\newcommand{\bt}{\beta}

\catcode`\@=12

\newcommand{\al}{\alpha}

\def\del          {\partial}

\newcommand{\reef}[1]{(\ref{#1})}
\renewcommand{\eqref}[1]{(\ref{#1})}
\def\ph1{\phantom{1}}

\newcommand{\beq}{\begin{equation}}
\newcommand{\eeq}{\end{equation}}
\newcommand{\ba}{\begin{aligned}}
\newcommand{\ea}{\end{aligned}}
\newcommand{\beqa}{\begin{eqnarray}}
\newcommand{\eeqa}{\end{eqnarray}}
\newcommand{\beqar}{\begin{eqnarray*}}
\newcommand{\eeqar}{\end{eqnarray*}}

\title{Renormalization group flow of entanglement entropy on spheres}

\author[a]{Omer Ben-Ami,}
\author[a]{Dean Carmi}
\author[b]{and Michael Smolkin}

\affiliation[a]{Raymond and Beverly Sackler Faculty of Exact Sciences School of Physics and Astronomy Tel-Aviv University, Ramat-Aviv 69978, Israel}
\affiliation[b]{Center for Theoretical Physics and Department of Physics,\\
 University of California, Berkeley, CA 94720, U.S.A. }

\emailAdd{omerben@post.tau.ac.il}
\emailAdd{carmidea@post.tau.ac.il}
\emailAdd{smolkinm@berkeley.edu}

\abstract{We explore entanglement entropy of a cap-like region for a generic quantum field theory residing in the Bunch-Davies vacuum on de Sitter space. Entanglement entropy in our setup is identical with the thermal entropy in the static patch of de Sitter, and we derive a simple relation between the vacuum expectation value of the energy-momentum tensor trace and the RG flow of entanglement entropy. In particular, renormalization of the bare couplings and logarithmic divergence of the entanglement entropy are interrelated in our setup. We confirm our findings by recovering known universal contributions for a free field theory deformed by a mass operator as well as obtain correct universal behaviour at the fixed points. Simple examples of entanglement entropy flows are elaborated in $d=2,3,4$. In three dimensions we find that while the renormalized entanglement entropy is stationary at the fixed points, it is not monotonic. We provide a computational evidence that the universal `area law' for a conformally coupled scalar is different from the known result in the literature, and argue that this difference survives in the limit of flat space. Finally, we carry out the spectral decomposition of entanglement entropy flow and discuss its application to the F-theorem.}

\begin{document}

\maketitle

\section{Introduction}

Entanglement entropy is an important tool in condensed matter physics \cite{KitPres05,LevinWen,BenjMul,Grover:2011fa}, quantum field theory\cite{Srednicki:1993im,Holzhey:1994we,Calabrese:2004eu,Casini:2009sr,Calabrese:2009qy} and quantum gravity \cite{Bombelli:1986rw,Callan:1994py,Ryu:2006bv, Ryu:2006ef,Hubeny:2007xt,Solodukhin:2011gn,Lewkowycz:2013nqa,Faulkner:2013ana,Faulkner:2013ica,Bousso:2014sda,Ben-Ami:2014gsa,Engelhardt:2014gca}. Nowadays this technique is in the spot light of nonperturbative studies of the structure of quantum field theories (QFT).  In particular, constraints imposed by $c$-theorems, which describe the irreversible nature of renormalization group (RG) flows between the fixed points, is one of the examples where entanglement entropy methods are applied \cite{Casini:2004bw,Myers:2010xs,Myers:2010tj,Myers:2012ed,Liu:2012eea,Solodukhin:2013yha,Liu:2013una,Banerjee:2014oaa,Solodukhin:2014dva,Kawano:2014moa,Balasubramanian:2014bfa,Casini:2014yca}.\footnote{See also \cite{Nozaki:2014hna,Cardy:2014jwa,Herzog:2014fra,Hung:2014npa,Faulkner:2014jva,Goykhman:2015sga,Huang:2015bna,Miao:2015iba} for recent progress in applying entanglement entropy techniques in a QFT context.}  These theorems spring directly from Zamolodchikov's proof \cite{Zamolodchikov:1986gt} of irreversibility of RG flows in two dimensions. In general dimension, however, the mechanism of irreversibility is still unclear, but Cardy's conjecture \cite{Cardy:1988cwa} suggests that this feature is inherent to any QFT for any even dimensional space-time. Recently, a striking proof was found in four dimensions \cite{Komargodski:2011vj}, see also \cite{Komargodski:2011xv,Luty:2012ww}, but its generalization to six and higher dimensions is not straightforward \cite{Elvang:2012st}. 

A natural extension of Cardy's conjecture for quantum field theories in an odd dimensional space-time was proposed in \cite{Myers:2010xs,Myers:2010tj}. Based on the holographic studies of entanglement entropy in general dimension, they suggested that the universal coefficient of entanglement entropy for a spherical region is decreasing along the RG trajectory.  At the fixed points of RG flows in even dimensions this coefficient is proportional to the central charge used by Cardy to formulate his conjecture \cite{Holzhey:1994we,solo,Casini:2014aia,Casini:2011kv} , while for odd dimensions it was shown in \cite{Casini:2011kv}, see also \cite{Dowker:2010yj}, that this coefficient is directly related to the $F$-theorem \cite{Jafferis:2011zi,Klebanov:2011gs}.  These observations indicate that entanglement entropy provides a useful framework to study $c$-theorems in general odd and even dimensions.

Important progress in understanding $c$-theorems in three dimensions was made in \cite{Casini:2012ei} where Casini and Huerta used strong subadditivity of entanglement entropy \cite{Lieb:1973cp} to prove the three-dimensional F-theorem for any unitary and Lorentz invariant QFT. There is not yet an alternative derivation. However, it is certainly important to understand the key insights of the proof in terms of conventional field theoretic methods, and one of our goals in this paper is to make a step in this direction.

Motivated by \cite{Rosenhaus:2014zza}, we think of entanglement entropy as a scalar functional of the field theory couplings and geometry of the setup. Imposing various symmetries inherent to the underlying QFT yields a set of Ward identities and RG flow equations satisfied by the entanglement entropy functional \cite{Rosenhaus:2014nha}. The resulting identities depend on various correlators involving insertions of the modular Hamiltonian, and therefore from computational perspective they are not really tractable since the modular Hamiltonian is not known and not local in general. However, for a special class of co-dimension two entangling surfaces which exhibit rotational symmetry in the transverse space, the modular Hamiltonian can be identified with the angular evolution operator around a given entangling surface \cite{Bis75,Bis76,Kabat:1994vj} while the corresponding Ward identities and RG flow equations can be written in terms of the standard correlation functions. 

Flat entangling plane in Minkowski space-time is probably the simplest realization of such a symmetric setup. It has been shown in \cite{Kabat:1994vj} that entanglement entropy in this case equals thermal entropy of the Rindler wedge. Perturbative studies of this setup are useful for understanding the structure of both the universal entanglement entropy  \cite{Rosenhaus:2014woa,Rosenhaus:2014ula,Rosenhaus:2014zza}, see also \cite{Banerjee:2011mg,Allais:2014ata,Mezei:2014zla}, and the modular Hamiltonian \cite{Lewkowycz:2014jia}. There is also a way to use it in order to set a precise relation between the entanglement entropy and renormalization of the Newton's constant \cite{Casini:2014yca}. However, this setup has no built-in scale, and its topology is trivial, which makes it hard to believe that one can learn much about the non-perturbative aspects of $c$-theorems in higher dimensions based on this example.\footnote{However, see \cite{Casini:2014yca} for an alternative derivation of Zamolodchikov's theorem using this setup.} 

Our proposal in section \ref{sec:setup} is to use a rotationally symmetric entangling surface for a field theory residing in the Bunch-Davies vacuum on de Sitter space.\footnote{For a CFT this construction was studied in \cite{Myers:2010tj} and holographically in \cite{Myers:2010xs,Myers:2010tj}, see also \cite{Maldacena:2012xp} and \cite{Fischler:2013fba} for various aspects of the entanglement entropy in de Sitter space.}  Rotational symmetry ensures simple structure for the modular Hamiltonian while the curvature of maximally symmetric geometry effectively sets the renormalization group scale.\footnote{In the system we study, the scale of entangling surface is also defined by the radius of de Sitter space, therefore there is only one global geometric scale.} Such a setup is scalable in the sense defined by \cite{Liu:2012eea}, and RG flow equation for entanglement entropy can be expressed in terms of two-point function of the energy-momentum tensor. A further simplification takes place after we notice in section \ref{sec:ward} that the RG flow of entanglement entropy has simple interrelation with the vacuum expectation value (vev) of the energy-momentum tensor trace. This leads us to a quite interesting conclusion that the logarithmic divergences associated with renormalization of the bare operators and bare parameters in the energy-momentum tensor on de Sitter space are simply related to the universal entanglement entropy. We illustrate this peculiarity in section \ref{free} using the example of massive free field theories, where renormalization of the cosmological constant must be done even in the absence of interactions.

Our construction has much in common with the Rindler space and with the observations made by \cite{Kabat:1994vj}. For example, entanglement entropy equals thermal entropy for any QFT in our setup. However, there are several distinctions which make it more interesting. Firstly, the Euclidean continuation of the causal domain of entangling region in de Sitter space  has non trivial topology in comparison to its counterpart in Minkowski space (sphere versus flat plane). Secondly, the curvature of de Sitter space sets up a characteristic temperature which manifests itself through the finite terms in entanglement entropy. Finally, one can explore an RG flow of entanglement entropy between the fixed points by varying the radius of de Sitter space. Given that de Sitter space is more realistic than Minkowski background, such RG flows are worth investigation.

In section \ref{free} we build on the advantages of our setup to study simple RG flows. In particular, we find that renormalized entanglement entropy defined in \cite{Liu:2012eea,Casini:2012ei} is not monotonic on a three dimensional de Sitter space. Of course, this observation does not contradict \cite{Casini:2012ei} since there is no reason to expect that RG flows on de Sitter and Minkowski space-times share the same merits. Furthermore, in contrast to the flat space results in \cite{Klebanov:2012va}, our findings in section \ref{free} indicate that the renormalized entanglement entropy is stationary at the fixed points on a sphere. This qualitative difference between the backgrounds can be attributed to the absence of infrared (IR) divergences in our setup, since sphere can be thought of as a finite box. It should be noted, however, that this result is in certain tension with \cite{Klebanov:2012va}, see also \cite{Dowker:2012rp}, where non-stationarity was found not only for a scalar field theory living in flat space, but also for a conformally coupled scalar on a sphere. This disagreement is closely related to another observation made in section \ref{free} where we argue that the universal `area law' for a massive conformally coupled scalar field on a sphere is different from its counterpart in flat space \cite{Hertzberg:2010uv,Huerta:2011qi,Lewkowycz:2012qr}. This discrepancy is not sensitive to the size of de Sitter radius which points in favour of controversial \cite{Solodukhin:1996jt,Hotta:1996cq,Solodukhin:2011gn,Nishioka:2014kpa,Lee:2014zaa,Herzog:2014fra,Dowker:2014zwa,Casini:2014yca} difference  between the universal entanglement entropies for minimally and conformally coupled scalar fields in {\it flat} space. Possible explanation of this phenomenon is given in \cite{MRS}, see also \cite{Larsen:1995ax,Rosenhaus:2014ula,Rosenhaus:2014zza}.

Finally, in section \ref{sec:spec} we use \cite{Osborn:1999az} to elaborate the spectral decomposition of entanglement entropy flow in general dimension. This decomposition is  determined by an integral over a positive weight function which represents the spin-0 part of the spectral representation for two point function of the energy-momentum tensor. In particular, the rate of change of entanglement entropy with respect to RG scale has definite sign, provided that the integral over the weight function converges. In three dimensions it suggests that the finite part of entanglement entropy may play a role of the $c$-function. However, this conclusion is too na\"ive. Indeed, possible logarithmic divergences of the entanglement entropy for a QFT in three dimensions \cite{Hung:2011ta,Rosenhaus:2014zza} is a clear sign of potential problems with convergence of the spectral integral. Moreover, finite part of entanglement entropy at the IR fixed point is contaminated by various mass scales which are remnants of RG running into the fixed point \cite{Casini:2014yca}. In the deep infrared these finite terms are part of UV physics, and therefore they should be subtracted to distill the genuine universal piece of entanglement entropy at the IR fixed point.

\section{Setup}
\labell{sec:setup}

Let us consider a generic QFT living on a $d$-dimensional background de Sitter space of radius $R$ - a submanifold of $(d+1)$-dimensional Minkowski space described by the hyperboloid of one sheet,
\be
 x_1^2+x_2^2+...+x_d^2-x_0^2=R^2 ~.
\ee
We will assume that the field theory resides in a unique Bunch-Davies vacuum state, also known as the Euclidean vacuum, which is invariant under all the isometries. In global coordinates, 
\bea
 x_0&=&R\sinh\Big({t\over R}\Big)~,
 \non
 x_i&=&R\cosh\Big({t\over R}\Big)\Omega_i ~, \quad i=1,..,d~,
\eea
with $\sum_i \Omega_i^2=1$, the metric on de Sitter space takes the form
\be
 ds^2=-dt^2+R^2\,\cosh^2\Big({t\over R}\Big)d\Omega_{d-1}^2~.
\ee

\begin{figure}
\begin{center}
\includegraphics[width=7cm,height=9cm]{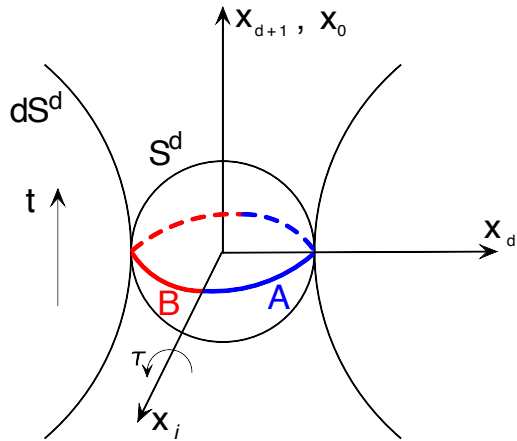}
\caption{Illustrating the setup of a QFT living on an $S^d$ sphere. The entangling surface (at $t=0$ and $\theta = \pi/2$) divides the equator into two equal cap regions (shown in blue and red).}
\label{dS}
\end{center}
\end{figure}

In what follows we aim to consider entanglement entropy of two equal cap-like regions $A$ and $B$ which are located on a $x_0=0$ ($t=0$) time slice of $dS^d$. This slice is invariant under $t\to -t$ which makes it possible to analytically continue the metric to the Euclidean signature, \ie $x_0\to ix_{d+1}$ or equivalently $t\to i\tau$. In particular, the problem of finding entanglement entropy boils down to studies of a QFT living on a $d$-dimensional sphere of radius $R$, see Fig.\ref{dS}.

By construction, the entangling surface, $\Sigma$, divides the equator of $S^{d}$ into two equivalent cap-like regions. In particular, it exhibits rotational symmetry in the transverse space which has topology of a 2-sphere. Hence, we find it convenient to foliate $S^d$ such that this symmetry becomes manifest,
\bea
 x_i&=&R\sin\theta\,\Omega_i \quad \text{for} \quad 1\leq i\leq d-1 \quad ,~0\leq\theta\leq{\pi\over 2}\quad, ~ \sum_{i=1}^{d-2}\Omega_i^2=1~,
 \non
 x_{d}&=&R\cos\theta\cos\tau ~,
 \non
 x_{d+1}&=& R\cos\theta\sin\tau~,
 \labell{coor}
\eea
where $(\tau,\theta)$ parametrize a two-dimensional transverse space to the entangling surface at $\theta={\pi\over 2}$.\footnote{In two dimensions $-{\pi\over 2}\leq\theta\leq{\pi\over 2}$, and entangling surface consists of two disjoint points at $\theta=\pm{\pi\over 2}$.} The infinitesimal line element on $S^d$ takes the form
\be
 ds^2=\sum_{i=1}^{d+1}dx_j^2=R^2\Big(\cos^2\theta d\tau^2+d\theta^2+\sin^2\theta d\Omega_{d-2}^2\Big)~.
 \labell{foliation}
\ee

It worth noting that the above foliation of $S^d$ corresponds to analytic continuation of the static patch of de Sitter to Euclidean signature, and it has been shown in \cite{Casini:2011kv} that if the field theory is conformal, then entanglement entropy for a sphere of radius R in Minkowski space is equivalent to the thermodynamic entropy of the thermal state in the static patch of de Sitter geometry. Of course, in the absence of conformal symmetry there is no simple relation between these entropies. However, by construction it is still true that  the entanglement entropy of two equal cap-like regions for a theory residing in the Bunch-Davies vacuum equals to the thermal entropy associated with a thermal state in the static patch of de Sitter. A quantitative manifestation of this (exact) relation is provided in section \ref{free}.

Now let us assume that the field theory on $S^d$ is some CFT deformed by a set of relevant operators $\mathcal{O}_i$ of scaling dimension $\Delta_i$ and associated  couplings $\lambda^i$. Since entangling surface exhibits rotational symmetry in the transverse space, the modular Hamiltonian is given by
\be
 K= - 2\pi\int_A T_{\mu\nu} \xi^\mu n^\nu + c' ~, \quad T_{\mu\nu}(x)={2\over\sqrt{g}} {\delta I \over \delta g^{\mu\nu}(x)} ~.
 \labell{entham}
\ee
where $I$ is the Euclidean action of the field theory, the integral runs over the region $A$ of the equator, $n^\nu=(R\cos\theta)^{-1}\del_{\tau}$ is normal to this region, $\xi^\mu=\del_{\tau}$ is the Killing vector associated with rotational symmetry around $\Sigma$ and $c'$ is an additive constant to ensure proper normalization of the reduced density matrix, \ie $\text{Tr}e^{-K}=1$.

Now recall the general flow equations that describe changes in the entanglement entropy under flow of the relevant couplings and deformation of the geometry \cite{Rosenhaus:2014nha},
\bea
\frac{\del S_{\mt{EE}}}{\del \lambda^i} &=&  - \int d^dx \sqrt{g(x)} \langle O_i(x) K\rangle_{\lambda} ~, \quad
 \mathcal{O}(x)  = {1\over \sqrt{g(x)}}{\delta I\over\delta\lambda^i(x)}~, \quad
\labell{relflow}
\\
 \frac{\delta S_{\mt{EE}}}{\delta g^{\mu \nu}(x)} &=& - \frac{\sqrt{g(x)}}{2} \langle T_{\mu \nu}(x) K\rangle_{\lambda} ~,
 \labell{flow}
\eea
where $\lambda$ collectively denotes all the couplings. These flows are directly related to the first law of entanglement \cite{Marolf:2003sq,Bhattacharya:2012mi,Blanco:2013joa,Wong:2013gua} \footnote{See also \cite{Allahbakhshi:2013rda} for a discussion of certain parallelism between the entanglement and the laws of thermodynamics.}. Of course, for a general deformation $K$ will not be given by \reef{entham} since rotational symmetry will be destroyed. However, if we restrict our consideration to a particular one parameter family of constant rescaling of the background, then rotational symmetry is preserved and \reef{flow} takes the following form
\be
 \int d^dx \, g^{\mu\nu}\frac{\delta S_{\mt{EE}}}{\delta g^{\mu \nu}(x)} =- \frac{1}{2} \int d^dx \, \sqrt{g(x)} \langle T(x) K\rangle_{\lambda} ~,
 \labell{gflow}
\ee
where $T(x)=g^{\mu\nu}T_{\mu\nu}(x)$ is the trace of the energy-momentum tensor, and integrals run over $S^d$.

The left hand side simplifies if we notice that for our choice of the foliation constant Weyl rescaling of the background metric is tantamount to a constant rescaling of the radius of $S^d$, and we obtain
\be
 R \frac{d S_{\mt{EE}}}{d R} =  \int d^dx \, \sqrt{g(x)} \langle T(x) K\rangle_{\lambda} ~.
 \labell{flow2}
\ee

Eq.\,\reef{flow2} can be also derived using the standard replica trick approach just because the rotational symmetry around the entangling surface makes it possible to introduce a well-defined conical defect, such that
\be
 S_\mt{EE}=\lim_{\epsilon \to 0} \( {\del\over \del\epsilon} +1 \) \log Z_{1-\epsilon} ~,
\ee
where $Z_{1-\epsilon}$ is the partition function on the $n$-fold cover of a sphere  and $\epsilon=1-n$. Hence \cite{Ryu:2006ef,Myers:2010tj},
\bea
 R \frac{d S_{\mt{EE}}}{d R} &=& - 2\lim_{\epsilon \to 0} \( {\del\over \del\epsilon} +1 \) \int d^dx \, g^{\mu\nu}(x) {\delta\over\delta g^{\mu\nu}(x)} \log Z_{1-\epsilon} 
 \non
 &=&  \lim_{\epsilon \to 0} \( {\del\over \del\epsilon} +1 \) \int d^dx \, \sqrt{g(x)} \langle T(x) \rangle_{1-\epsilon}
 \non
 &=&\int d^dx \, \sqrt{g(x)} \langle T(x) K \rangle ~,
\eea
where the last equality rests on \cite{Smolkin:2014hba,Hung:2014npa}.

Unfortunately, \reef{flow2} is ambiguous when supports of the energy-momentum tensor and the modular Hamiltonian collide. Hence, we resort to the Ward identities which help  to circumvent this ambiguity.

\section{Ward identities}
\label{sec:ward}

In this section we use the Ward and trace identities to constrain correlation functions involving the modular Hamiltonian and the energy-momentum tensor. They will help us to fix certain ambiguities involving situations when the energy-momentum tensor collides with the support of modular Hamiltonian. 

Let us consider the vacuum entanglement entropy of a generic surface $\Sigma$ for a quantum field theory defined on an arbitrary curved space. Assuming that regularization preserves invariance under the diffeomorphisms $g_{\mu\nu}\to g_{\mu\nu}+\nabla_{(\mu} v_{\nu)}$, where $\nabla_\mu$ and $v_\mu$ stand for a covariant derivative and arbitrary vector field respectively, and taking into account that entanglement entropy is a scalar functional, implies\footnote{The most general Ward identity of this type, which also accounts for the possibility of local sources $\lambda^i(x)$, takes the form
\be
 \int d^dx \( -(\nabla^\mu v^\nu+\nabla^\nu v^\mu){\delta\over \delta g^{\mu\nu}(x)} + v^\mu\del_\mu \lambda^i {\delta\over\delta \lambda^i} \)S_{\mt{EE}} =0 ~,
\ee
or equivalently
\be
  \nabla^\mu \langle T_{\mu\nu}(x) K \rangle = \del_\mu\lambda^i\langle O_i(x) K\rangle~.
\ee
Ultimately, we will be interested in $\lambda^i(x)=\lambda^i$ the physical coupling constants, and therefore we drop terms involving $\del_\mu\lambda^i$.
} 
\be
 0=-\int d^dx (\nabla^\mu v^\nu+\nabla^\nu v^\mu){\delta\over \delta g^{\mu\nu}(x)} S_{\mt{EE}} = - \int d^dx \, \sqrt{g} \, v^\nu \nabla^\mu \langle T_{\mu\nu}(x) K \rangle ~,
\ee
where in the second equality we substituted the definition $S_{\mt{EE}}=-\text{Tr}\hat\rho\log\hat\rho=\langle K\rangle$ and used normalization of the density matrix to drop $\langle \delta K/\delta g^{\mu\nu}\rangle=0$, see \cite{Rosenhaus:2014zza}. Hence, we conclude
\be
 \nabla^\mu \langle T_{\mu\nu}(x) K \rangle =0 ~.
 \labell{delTK}
\ee

Let us contrast this formula with the standard Ward identity for the two point function of the energy-momentum tensor. Starting from the counterpart of \reef{delTK},
\be
 \nabla^\mu \langle T_{\mu\nu}(x)\rangle =0 ~,
 \labell{delT}
\ee
which one obtains from the condition that the effective action, $W\equiv\log Z$ with $Z$ being a partition function, is a scalar functional under diffeomorphisms, and differentiating it with respect to the metric, yields \cite{Osborn:1999az}
\be
\nabla^\mu \langle T_{\mu\nu}(x) T_{\al\bt}(y)\rangle = \nabla_\nu \(\delta^\sigma_\alpha \delta^\rho_\bt \, \delta^d(x,y) \) \langle T_{\sigma\rho}(x) \rangle
+2\nabla_\sigma\(  \delta^\sigma_{(\alpha} \delta^\rho_{\bt)} \, \delta^d(x,y) \langle T_{\rho\nu} (x) \rangle \) ~,
 \labell{delTT}
\ee
where $\delta^d(x,y)=\delta^d(x-y)/\sqrt{g(x)}$,whereas the 2-point function is defined by
\bea
  \sqrt{g(x)}\sqrt{g(y)} \, \langle T_{\mu\nu}(x)T_{\al\bt}(y)\rangle&\equiv& 
 4 \,  {\delta^2 W\over \delta g^{\mu\nu} (x) \delta g^{\al\bt} (y)}  ~.
 \labell{2pT}
\eea

The right hand side of \reef{delTT} vanishes up to contact terms involving $\delta$-functions, while the right hand side of \reef{delTK} vanishes identically. In particular, in the special case when the modular Hamiltonian is local and is given by \reef{entham}, it would be na\"{\i}ve to use \reef{2pT} to evaluate the right hand side of \reef{flow2}. One needs to modify this correlator in the limit of coincident points such that \reef{delTK} holds identically. A necessary modification in a slightly different context was carried out in \cite{Osborn:1999az}, and we review it here.  

One starts from noting that the vacuum expectation value of any local scalar operator on a sphere is just a constant and also
\be
 \langle T_{\mu\nu} (x) \rangle = -{1\over d} \, {C\over R^d} \, g_{\mu\nu}(x)~,
 \labell{1pT}
\ee
where $C$ is some dimensionless function of the couplings. Eq.~\reef{delT} is trivially satisfied, whereas \reef{delTT} becomes
\be
 \nabla^\mu \langle T_{\mu\nu}(x) T_{\al\bt}(y)\rangle = -{1\over d} \, {C\over R^d}\( \nabla_\nu \delta^d(x,y) g_{\al\bt} (y) 
 +2 \nabla_\sigma\big(  \delta^\sigma_{(\alpha} \delta^\rho_{\bt)} \, \delta^d(x,y)g_{\rho\nu}(x)\big) \)
\ee
Next define,
\be
  \langle T_{\mu\nu}(x) T_{\al\bt}(y)\rangle_{\text{con}}=  \langle T_{\mu\nu}(x) T_{\al\bt}(y)\rangle 
 + {1\over d} \, {C\over R^d}\big( g_{\mu\al} g_{\nu\bt} + g_{\mu\bt} g_{\nu\al} + g_{\mu\nu} g_{\al\bt} \big) \delta^d(x,y) ~.
 \labell{conTT}
\ee
where $\langle \ \rangle_{\text{con}}$ means ``conserved", as in \cite{Osborn:1999az}. This 2-point correlator satisfies a desired conservation equation \cite{Osborn:1999az},
\be
 \nabla^\mu \langle T_{\mu\nu}(x) T_{\al\bt}(y)\rangle_{\text{con}}=0 ~,
\ee
and therefore it should be used in \reef{flow2} to ensure that the Ward identity \reef{delTK} is satisfied. 

Now recall that for our setup homogeneous Weyl rescaling obeys
\be
 2\int d^dx g^{\mu\nu}(x) {\delta \over \delta g^{\mu\nu}(x)} = -R {d\over dR}~.
 \labell{equiv}
\ee
In particular, applying this operator to the both sides of \reef{1pT} leads to
\be
 -\int d^d y\sqrt{g(y)} \langle T(y) T_{\mu\nu}(x)\rangle + d \langle T_{\mu\nu} (x) \rangle =  {1\over d} \, R{d\over dR} \({C\over R^d}\) \, g_{\mu\nu}(x) 
 - {2 \over d} \,  {C\over R^d} \, g_{\mu\nu}(x) ~.
\ee
Or equivalently, using \reef{conTT}
\be
 -\int d^d y\sqrt{g(y)} \langle T(y) T_{\mu\nu}(x)\rangle_{\text{con}} =  {g_{\mu\nu}(x) \over d} \, R{d\over dR} \({C\over R^d}\)  ~.
\ee
Substituting this formula into \reef{flow2} and using \reef{entham}, we obtain one of our main results\footnote{We stress that unlike the one point function $\left<K\right>$, the (divergent) constant $c'$ drops out of connected correlators. 
}
\be
R\frac{d S_{\mt{EE}}}{d R} =  {2\pi\over d} \, R{d\over dR} \({C\over R^d}\)  \int_A \xi\cdot n = { \Omega_d R^{d+1} \over d} \,{d\over dR} \({C\over R^d}\)~,
\labell{dSdR}
\ee
where $\Omega_d$ is the surface area of a unit $d$-dimensional sphere,
\be
\Omega_d={2\pi^{d+1\over 2}\over \Gamma\({d+1\over 2}\)}~.
\ee

It worth mentioning that \reef{dSdR} relates finite quantities on both sides of the equation since $C$ is finite by definition. In particular, this formula eliminates the power law divergent terms of EE such as the well known `area law'.  Of course, there is nothing bad about absence of these terms since they are scheme dependent and therefore vanish under appropriate choice of the regularization scheme (\eg dimensional regularization). However, the logarithmic divergence of EE is always retained, and its coefficient is scheme independent. This observation may cast doubts on possible applications of \reef{dSdR} for a QFT in even dimensions, where the universal EE is associated with a logarithmic divergence. In section \ref{free} we elaborate on simple examples which clarify this subtlety. We show that the universal EE in our setup is directly related to the logarithmic divergences of the bare parameters defining the theory, and the standard renormalization results in a finite EE with the universal data being encoded in the coefficient of the (finite) logarithmic running. Throughout the paper we apply dimensional regularization scheme, and therefore non-universal divergences of EE will be absent from our calculations.

\subsection{Thermal interpretation}

As pointed out in section \ref{sec:setup}, entanglement entropy equals thermal entropy in the system under study, $S_\mt{EE}=S_\mt{Th}$. Therefore \reef{dSdR} should have a simple interpretation in terms of standard thermodynamics. We provide such an argument below, see also \cite{Casini:2011kv} for thermal analysis of $S_\mt{EE}=S_\mt{Th}$ at the fixed point.

Let us start from the following thermodynamic relation,
\begin{equation}
\label{eq:Sthermal}
S_\mt{Th}=\beta (U-F) ~,
\end{equation} 
where $U$ and $F$ are thermal and free energies respectively, and $\beta=2\pi R$ is the inverse temperature. As usual, $U$ is given by the expectation value of the operator which generates translations around the thermal circle parametrized by $\tau$ in \reef{foliation}. Hence, based on \reef{1pT} %we get\cite{Casini:2011kv}:
\begin{equation}
\beta U=\frac{\Omega_d }{d}\, C~.
\end{equation}
Now we can operate with $R {d\over dR}$ on both sides of \reef{eq:Sthermal}. Using \reef{equiv}, our convention \reef{entham} for the energy-momentum tensor, and recalling that $\beta F= - W$, yields 
\begin{equation}
R {d\over dR}(\beta F)=\Omega_d \, C \quad \Rightarrow \quad R {d\over dR}S_\mt{Th}={\Omega_d\over d} \left(R {d C\over dR} - d\, C \right)~.
\end{equation}
This is exactly \reef{dSdR}, thus we showed that $S'_{EE} = S'_{Th}$. Since at the fixed point we have $S_{EE} = S_{Th}$, we conlcude that this is true also outside the fixed point.

\subsection{RG equation for entanglement entropy}

The renormalization group running of entanglement entropy in our setup can be readily evaluated. Based on \reef{1pT}, we have
\be
 {\del C\over \del \lambda^i}= - R^d {\del \over \del \lambda^i}\langle T(x) \rangle =  R^d \int d^dy \sqrt{g(y)} \langle O_i(y) T(x)  \rangle + d R^d \langle O_i(x)\rangle
 =R{d\over dR} \( R^d \langle O_i(x) \rangle \)~,
\ee
where in the last equality we used homogeneity of the sphere to replace integral over $y$ with integral over $x$ and then applied the correspondence \reef{equiv}. Next we combine this result with the trace Ward identity\footnote{For brevity we suppress index $i$ of the relevant couplings. Sum over this index is implicitly assumed.} \cite{Osborn:1993cr,Osborn:1999az}
\be
 \langle T \rangle+(d - \Delta - \beta) \lambda \langle O \rangle = \mathcal{A} \quad \Rightarrow \quad 
 C= R^d \Big(  (d - \Delta - \beta)\lambda\langle O \rangle -\mathcal{A}  \Big)~,
\ee
where $\beta$ is the beta function, and $\mathcal{A}$ is the trace anomaly formed from the Riemann tensor and its derivatives, \ie  $\mathcal{A}=0$ for any odd dimensional space-time while for even dimensional sphere $\mathcal{A}\propto R^{-d}$. As a result, we obtain \cite{Osborn:1999az}
\be
 \( R{d \over dR} - (d - \Delta - \beta)\lambda {\del\over\del\lambda} \) C = - R{d \over dR} \(R^d \mathcal{A} \) =0 ~.
\ee
This equation combined with \reef{relflow}, \reef{1pT}, \reef{dSdR} and the definition of the modular Hamiltonian \reef{entham} leads to $\lambda\frac{\partial S}{\partial \lambda}=-\frac{1}{d}\Omega_d\lambda\frac{\partial C}{\partial \lambda} + \Omega_dR^d\left<O(x)\right>$,\footnote{When deriving this relation it should be noted that the difference between $\lambda\frac{\partial C}{\partial\lambda}$ and $\lambda\frac{\partial S}{\partial\lambda}$ stems from the fact that $\left<\frac{\del T_{\mu\nu}}{\del \lambda}\right>\sim g_{\mu\nu}\left<O\right>$ but $\left<\frac{\del K}{\del \lambda}\right>=0$, as can be seen from the requirement $\text{Tr}\,e^{-K}=1$} substituting this we get:
\be
 \( R{d \over dR} - (d - \Delta - \beta)\lambda {\del\over\del\lambda} \) S_{\mt{EE}} = \Omega_d \, R^d \,\mathcal{A} ~.
 \labell{flow3}
\ee

Given that $\mathcal{A}$ is constant on a sphere, we deduce that the right hand side is just the integrated trace anomaly. This equation is a particular realization of the general idea presented in \cite{Rosenhaus:2014nha}. Here, however, we keep the finite anomalous term on the right hand side of \reef{flow3} since we are interested to evaluate finite logarithms which combine with the universal divergence of entanglement entropy to form a dimensionless term.

\subsection{Conformal fixed point}
\label{CFT}

Let us consider a rotationally symmetric entangling surface for a CFT residing in a vacuum state on some curved manifold.\footnote{It should not necessarily be our setup, \eg waveguide geometry \cite{Hertzberg:2010uv,Lewkowycz:2012qr} works equally well.} In this case, the right hand side of \reef{gflow} is completely fixed by the trace anomaly
\be
 \langle T(x) \rangle = \A=\sum_n b_n I_n(x) -2 (-1)^{d\over 2}a E_d(x)+B'\nabla_\mu J^\mu(x)~,
 \labell{trace}
\ee
which defines the central charges for a CFT in an even number of
dimensions. Each term on the right-hand side is constructed from the background geometry. $I_n$ are Weyl invariant combinations of the Weyl tensor, the Cotton tensor and the Bach tensor as well as their covariant derivatives. These basis tensors all vanish on any conformally flat background. The last term in eq.~\reef{trace} is a scheme-dependent total derivative which can be eliminated by adding a covariant counter-term to the effective action. Finally, $E_d$ is the Euler density in $d$ dimensions,
\be
E_{2p}(\cR) \equiv \frac{1}{ (8\pi)^p\, \Gamma(p+1)}\ \delta_{\mu_1\,\mu_2\,\cdots\,
\mu_{2p-1}\,\mu_{2p}}^{\nu_1\,\nu_2\,\cdots\, \nu_{2p-1}\,\nu_{2p}}\
\cR^{\mu_1\mu_2}{}_{\nu_1\nu_2}\,\cdots\,
\cR^{\mu_{2p-1}\mu_{2p}}{}_{\nu_{2p-1}\nu_{2p}}\,,
 \labell{term0}
 \ee
where $\delta_{\mu_1\,\mu_2\,\cdots\,
\mu_{2p-1}\,\mu_{2p}}^{\nu_1\,\nu_2\,\cdots\, \nu_{2p-1}\,\nu_{2p}}$ 
denotes a totally antisymmetric product of $2p$ Kronecker delta
symbols and the normalization ensures that $\int_{S^d}d^d\!x\sqrt{g}\, E_d =2$. 

Varying \reef{trace} with respect to $g^{\mu\nu}(y)$ and using the definition \reef{2pT}, yields
\be
 \langle T(x) T_{\mu\nu}(y)\rangle - \delta^d(x,y) g_{\mu\nu}(x) \langle T(x)\rangle - 2 \delta^d(x,y) \langle T_{\mu\nu}(x)\rangle =   - {2\over\sqrt{g(y)}} {\delta\A(x)\over\delta g^{\mu\nu}(y)}~.
\ee
In our setup the vev of the energy-momentum tensor takes a simple form \reef{1pT}, and therefore according to the definition of conserved 2-point function \reef{conTT}, we may identify the left hand side with $\langle T(x) T_{\mu\nu}(y) \rangle_{\text{con}}$. We will assume that similar redefinition exists for any system with rotationally invariant entangling surface.\footnote{It would be interesting to work out the details of this redefinition based on the requirement that Ward identity \reef{delTK} holds. Our assumption here rests on the observation that in the replica trick approach there are no $\delta$-functions away from the entangling surface.} Hence, \reef{gflow} at the fixed point reads
\be
  \int d^dx \, g^{\mu\nu}\frac{\delta S_{\mt{EE}}}{\delta g^{\mu \nu}(x)}
   = - 2\pi \int d^dx \sqrt{g(x)} \int d^{d-1}y \sqrt{h(y)}
 \({1\over\sqrt{g(y)}} {\delta\A(x)\over\delta g^{\mu\nu}(y)}\)\xi^\mu n^\nu~, 
\ee
where $h_{\mu\nu}$ denotes the induced metric on a region $A$ enclosed by the entangling surface. 

The right hand side can be readily evaluated in our case. Indeed, $I_n$ terms in the trace anomaly \reef{trace} play no role since they are at least quadratic in the building blocks which vanish for a conformally flat background. Total derivatives can be ignored,\footnote{Variation of the total derivative is given by
\be
 \delta(\nabla_\mu J^\mu) =(\delta\nabla_\mu)J^\mu+\nabla_\mu (\delta J^\mu) ~.
\ee
First term on the right hand side vanishes since $J^\mu=0$ on a sphere, whereas the integral of the last term vanishes since sphere has no boundaries. Therefore we conclude that the net contribution of the total derivatives to entanglement entropy flow is zero.
}
whereas the contribution of the Euler density is easy to evaluate since its integral is a topological term, \ie
\be
0={\delta \over\delta g^{\mu\nu}(y)}  \int d^dx \sqrt{g(x)} E_d(x) =  \int d^dx\sqrt{g(x)} \( {\delta E_d(x)\over\delta g^{\mu\nu}(y)} 
- {1\over 2}  g_{\mu\nu}(x) E_d(x) \delta(x-y)\)~.
\ee
Hence,
\be
 \int d^dx\sqrt{g(x)} {\delta E_d(x)\over\delta g^{\mu\nu}(y)}=  {\sqrt{g(y)} \over 2} \, g_{\mu\nu}(y)  \, E_d(y) ~.
\ee
Now $E_d$ is obviously constant on a sphere, therefore we can use our choice of normalization condition to get
\be
E_d |_{S^d}={2\over \Omega_d}R^{-d} ~.
\ee
Combining, yields
\be
 R \frac{d S_{\mt{EE}}}{d R} = {8\pi (-1)^{{d\over 2}-1} a \over \Omega_d \, R^d} \int d^{d-1}y \sqrt{h(y)} ~ \xi \cdot n =4 (-1)^{{d\over 2}-1} a ~.
 \labell{EEanom}
\ee
In accord with \cite{Myers:2010xs,Myers:2010tj}, see also \cite{Casini:2011kv}.

\section{Free fields on sphere}
\labell{free}

In this section we use free massive fields to elaborate on various properties of the formalism that we have developed in the previous section.

\subsection{Conformally coupled scalar}

Let us consider a free massive scalar field on a $d$-dimensional sphere of radius $R$
\be
I=\int_{S^d} \( {1\over 2}(\del\phi)^2 + {1\over 2} m^2\phi^2 + {1\over 2} \xi_c \cR \, \phi^2 \)~,
\labell{freeS}
\ee
where $\xi_c={d-2\over 4(d-1)}$ is the conformal coupling, $\cR={d(d-1)\over R^2}$ is the Ricci scalar of a sphere and 
$m^2$ is the mass of the field. The energy-momentum tensor is given by
\be
T_{\mu\nu}=\nabla_\mu\phi\nabla_\nu\phi - g_{\mu\nu} \( {1\over 2}(\del\phi)^2 + {1\over 2} m^2\phi^2 + {1\over 2} \xi_c \cR \, \phi^2 \) 
+\xi_c\cR_{\mu\nu}\phi^2 + \xi_c\(g_{\mu\nu}\nabla^2-\nabla_\mu\nabla_\nu\)\phi^2~.
\ee
Hence,
\be
T=g^{\mu\nu}T_{\mu\nu}=-m^2\phi^2 -{d-2\over 2} \phi(-\nabla^2+\xi_c\cR+m^2)\phi~.
\labell{imptr}
\ee
Discarding equation of motion operator leads to 
\be
C_\phi=m^2\,R^d \langle\phi^2\rangle ~.
\labell{Cphi}
\ee

The vev of $\phi^2$ is given by the coincident point limit of the Green's function which solves the Green's equation on $S^d$ 
\be
\[-{1 \over R^2 \sin^{d-1}\chi}
{\partial \over \partial \chi}\(\sin^{d-1}\chi{\partial \over \partial \chi}\)+\xi_c\cR+m^2\]
G_m(\chi) =\delta^{d}\(R \, \chi \)
\quad ,
\label{green2}
\ee
where $\chi$ is the polar angle on a sphere, and we used rotational symmetry to bring one of the points to the north pole. 

To solve the above equation, one needs to impose regularity at $\chi=\pi$, and 
\be
G_m(\chi) \sim {(R\chi)^{2-d}\over (d-2)\Omega_{d-1}} \quad \text{for} \quad \chi\ll 1~.
\ee
which corresponds to a scalar potential created by a unit charge placed at $\chi=0$. The general solution which satisfies these conditions is 
\bea
G_m(\chi)&=&{R^{2-d}\over (4\pi)^{d/2}} {\Gamma(\lambda)\Gamma(-\lambda+d-1)\over\Gamma(d/2)}  \ _{2}F_1(\lambda~,~ d-1-\lambda~;~{d\over 2}~;~\cos^2{\chi\over 2})~,
\labell{greensol}
\\
\lambda&=&{d-1\over 2}+i\sqrt{(mR)^2 - {1\over 4}}~.
\eea
Taking the limit $\chi\to 0$ and dropping a mass independent (divergent) constant\footnote{The mass independent term behaves as $\chi^{2-d}$, and therefore it diverges as power law in $d>2$. Such divergences are scheme dependent, and therefore one can choose a particular scheme where it vanishes, \eg in dimensional regularization one gets $\chi^{2-d}\to 0$ for $d<2$, and therefore analytic continuation to higher $d$ also vanishes.}, we have 
\be
\langle \phi^2 \rangle = {\Gamma(1-d/2)\Gamma(\lambda)\Gamma(d-1-\lambda)\over \pi (4\pi)^{d/2} R^{d-2}} \sin\big({\pi\over 2}(d-2\lambda)\big) ~.
\labell{vacphi2}
\ee
In particular, $\langle \phi^2 \rangle$ is finite for odd $d$, whereas for even $d$ it exhibits a simple pole which corresponds to the logarithmic divergence. We treat these cases separately.

\subsubsection*{Odd $d$}
For odd $d$, we have
\bea
C_\phi&=& {\pi (mR)^2 \coth(\pi\sqrt{m^2R^2-1/4})\over \sqrt{m^2R^2-1/4}}
{ (-)^{d-1 \over 2} \over (4\pi)^{d/2} \Gamma\({d\over 2}\) } \prod_{j=1}^{d-1\over 2} \( (d/2-1/2-j)^2-1/4+m^2R^2\)
\non
&=&{\pi (mR)^2 \coth(\pi\sqrt{m^2R^2-1/4})\over \sqrt{m^2R^2-1/4}}
{ (-)^{d-1 \over 2} \over (4\pi)^{d/2} \Gamma\({d\over 2}\) }  
\\
&\times& \( (mR)^{d-1}+{(d-1)(d^2-5d+3)\over 24}(mR)^{d-3} 
+\ldots+{\pi\over 2\,\Gamma\({2-d\over 2}\)\Gamma\({4-d\over 2}\)}\)~.
\nonumber
\eea
Appearance of the hyperbolic function in both the $C_\phi$ and the entanglement entropy after using \reef{dSdR} is not surprising. Indeed, such functions are the direct associates of any thermal computation, whereas the state in de Sitter space has thermal interpretation. To get rid of the thermal effects, let us consider the behaviour of entanglement entropy in the IR limit, $mR\gg 1$. In this regime curvature corrections are negligibly small, while thermal effects are exponentially suppressed, and we get
\begin{flalign}
C_\phi &\underset{\mt{$mR\gg 1$}}{=}
{ (-)^{d-1 \over 2} \pi \over (4\pi)^{d/2} \Gamma\({d\over 2}\) }  \( (mR)^d+{d(d-2)(d-4)\over 24}(mR)^{d-2} 
+\ldots\)\\
R{d S_\mt{univ}^\mt{scalar}\over dR} &\underset{\mt{$mR\gg 1$}}{=} {(d-2)(d-4)\over 24(d-1)}{ (-)^{d+1 \over 2} \pi \over (4\pi)^{d-2\over 2} \Gamma\({d\over 2}\) } A_\Sigma m^{d-2} + \ldots
\end{flalign}
where ellipsis encapsulate curvature corrections to the leading order term also known as universal `area law'. As shown in Appendix \ref{freeflat}, this term is identical to the universal `area law' of entanglement entropy of a half space for a conformally coupled scalar field residing in the Minkowski vacuum.  This result is a consequence of the fact that any surface and any background are locally flat.
\subsubsection*{Renormalized Entanglement Entropy in 3D}
For a vacuum state in 3-dimensional flat space QFT, the general pattern for EE of a disk is
\begin{eqnarray}
S_\mt{EE} = c_1{R\over \delta} - c_0 ~,
\end{eqnarray}
where $R$ is the radius of the disk and $\delta$ is a UV cut off. At the fixed point, $c_0$ and $c_1$ are some constants,\footnote{At the IR fixed point there could be additional terms, which are remnants of the RG flow \cite{Casini:2014yca}, \eg the universal `area law' $\sim\, mR$ for massive QFTs \cite{Hertzberg:2010uv,Huerta:2011qi,Lewkowycz:2012qr}. However, the characteristic (relevant) scale of such terms is very large in the deep IR, and therefore in this regime they are not really distinguishable from the $R/\delta$ term. Hence, we do not write them out. } whereas outside the fixed point $c_0$ is a function of $R$ and various scales characterizing a given QFT. Obviously, $c_1$ is  scheme dependent while $c_0$ is universal, therefore the authors of \cite{Liu:2012eea,Casini:2012ei,Liu:2013una} defined the so-called renormalized entanglement entropy (REE) to isolate the universal contribution,\footnote{See \cite{Liu:2012eea,Liu:2013una} for definition of REE in general dimension.}
\begin{eqnarray}
\mathcal{S}_3 \equiv R\frac{dS_\mt{EE}}{dR} - S_\mt{EE} \equiv RS_\mt{EE}'-S_\mt{EE}~.
\end{eqnarray}

Using strong sub-additivity of EE \cite{Lieb:1973cp}, Casini and Huerta proved that $S_\mt{EE}'' \leq 0$ for a disk in flat space. Hence, $\mathcal{S}_3'  = RS_\mt{EE}''\leq 0$, which clearly indicates that REE has a monotonic RG flow in flat space, and \cite{Casini:2012ei}
\begin{eqnarray}
\Delta c_0 = c_0^\mt{UV} - c_0^\mt{IR} = - \int_0^\infty dR \,R \, S_\mt{EE}''\geq 0 ~.
\label{Fthm}
\end{eqnarray}

Of course, the RG flow itself is ambiguous but the fixed points satisfy an F-theorem \cite{Jafferis:2011zi,Klebanov:2011gs}. Furthermore, conformal symmetry relates the REE at the fixed points in flat space to the universal EE in our setup \cite{Casini:2011kv}, hence it is worthwhile to explore REE on a sphere outside the fixed points. 

\begin{figure}[t]
	\centering
	\includegraphics[width=0.7\linewidth]{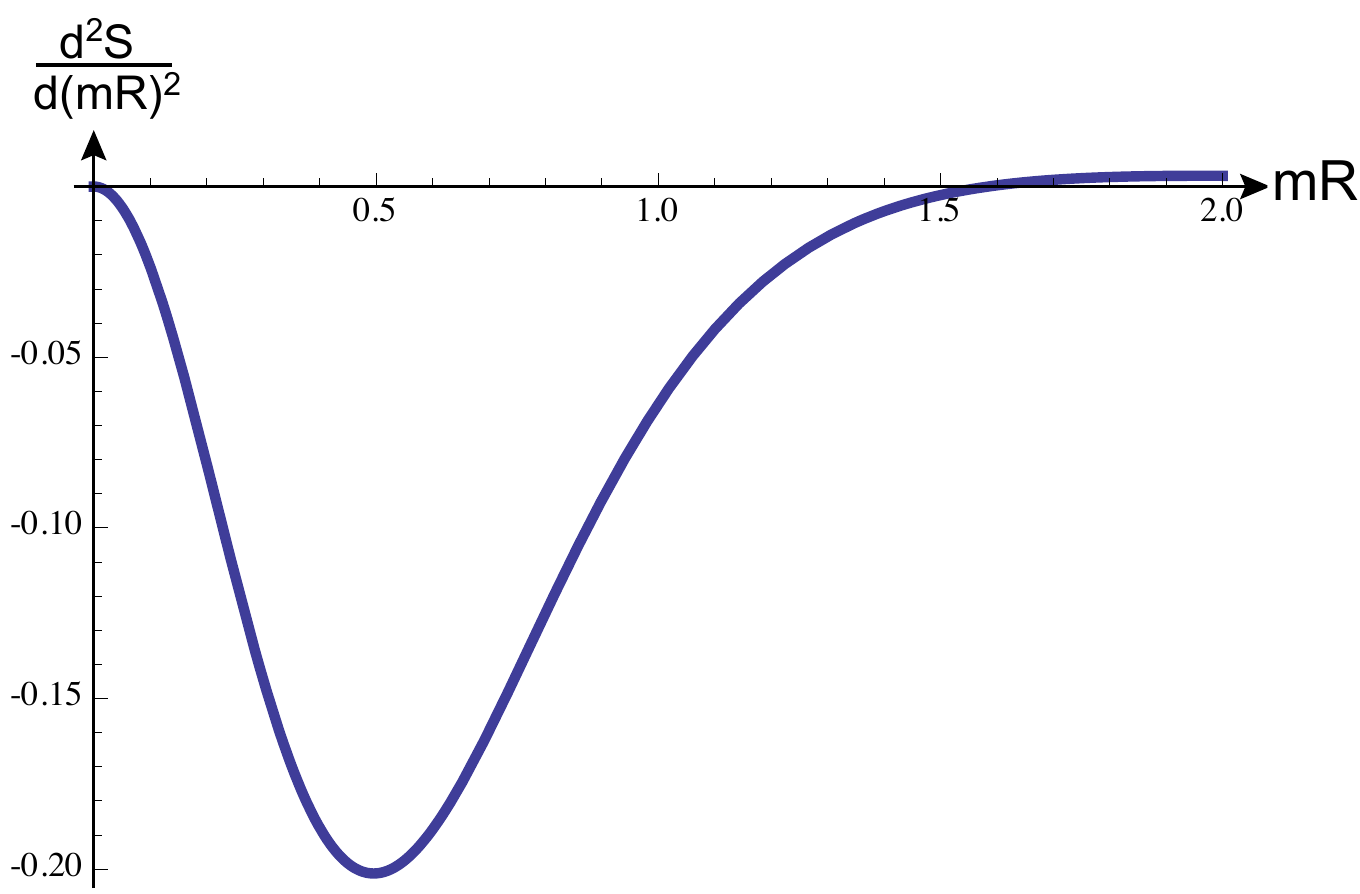}
	\caption{$\frac{d^2S_\mt{EE}}{d(mR)^2}$ for a massive conformally coupled scalar field on a 3-sphere. Unlike its counterpart in Minkowski space, $\frac{d^2S_\mt{EE}}{d(mR)^2}$ changes sign around $mR\sim 1.6$. }
	\label{fig:S''forscalarin3d}
\end{figure}

For a conformally coupled scalar field, $S''_\mt{EE}$ can be readily evaluated based on \reef{dSdR}, \reef{Cphi} and \reef{vacphi2}. Fig. \ref{fig:S''forscalarin3d} shows the corresponding plot, and it can be seen that $S''_\mt{EE}$ \underline{changes} sign. A similar issue is discussed in \cite{Klebanov:2011gs} for the free energy on a sphere, where the authors argue that additional subtractions are necessary because of emergence of the cosmological constant term in the IR limit. Although REE is not monotonic, other choices of subtraction scheme might result in a monotonic flow as was suggested for the free energy in \cite{Klebanov:2011gs}.

Moreover, REE in our setup exhibits stationarity at the fixed points, \ie $S''_\mt{EE}\to0$ as $mR\rightarrow0,\,\infty$. This result should be contrasted with numerical studies in \cite{Klebanov:2012va}, where it has been shown that REE for a disk in flat space is not stationary in the massless limit.  Similar considerations for the free energy showed that although the IR divergent free energy obeys stationarity, the existence of a subtraction scheme that maintains stationarity, monotonicity and analyticity is not obvious \cite{Klebanov:2011gs}.      

We have calculated $\Delta c_0\sim 0.0638$ which agrees with the expected result from the free energy calculation of a conformally coupled massless scalar in the UV and an empty theory in the IR, in spite of thermal effects mentioned above.

\subsubsection*{Even $d$}

For even $d_0$ the pole structure of $C_\phi$ is given by
\begin{flalign}
C_\phi&=(-)^{d_0/2}\,m^2R^2\,{\prod_{j=1}^{(d_0-2)/2} \( (d_0/2-1/2-j)^2+m^2R^2-1/4\)  \over (4\pi)^{d_0/2} \Gamma\({d_0\over 2}\)} {2\over d-d_0} + \ldots
\non
&={ (-)^{d_0/2} \over (4\pi)^{d_0/2} \Gamma\({d_0\over 2}\)} \((mR)^{d_0} + {d_0(d_0-2)(d_0-4)\over 24} (mR)^{d_0-2}  \right.\\
&\left.+\frac{d_0 (d_0-2) (d_0-4) (d_0-6) \left(5 d_0^2-18 d_0+4\right)}{5760}(mR)^{d_0-4}+ \ldots 
+ 2\, {\Gamma^2\big({d_0\over 2}\big) \over d_0-2} (mR)^4 \){2\over d-d_0}+\ldots 
\nonumber
\end{flalign}
The universal divergence of the entanglement entropy is obtained by substituting this expression into\footnote{As usual, simple poles in the dimensional regularization scheme correspond to logarithmic divergences, and we use the following dictionary $\log(m\delta)={1\over d-d_0}$, see Appendix \ref{freeflat}.} \reef{dSdR}
\bea
&&R {dS_\mt{univ}^\mt{scalar} \over dR}= {(d_0-2)(d_0-4)\over 12(d-1)} { (-)^{d_0/2+1}  A_{\Sigma} \over  (4\pi)^{(d_0-2)/2} \Gamma\({d_0\over 2}\)} 
\labell{Sphi}
\\
&&\times\(  m^{d_0-2} + \frac{(d_0-6) \left(5 d_0^2-18 d_0+4\right)}{120}\frac{m^{d_0-4}}{R^{2}} 
%\non
%&&\left.
+ \ldots + 24 {\Gamma^2\big({d_0\over 2}\big) \over d_0(d_0-2)^2 } \, {m^4 \over R^{d_0-6}} \)\log(m\delta) ~.
\nonumber
\eea

The leading order term in the limit $mR\gg1$ represents the universal `area law'. In this limit curvature effects can be disregarded, and the system can be approximated by a free scalar field theory living in the Rindler wedge \cite{Bianchi:2012ev}, therefore it should be possible to recover the same result by considering the entanglement entropy of a half space for a massive scalar field residing in the Minkowski vacuum state \cite{Kabat:1994vj}. We carry out this computation in Appendix \ref{freeflat} and find full agreement. However, the resulting expression for the universal `area law' is not the same as in, \eg  \cite{Hertzberg:2010uv,Huerta:2011qi,Lewkowycz:2012qr}. Possible interpretation of this discrepancy is given in \cite{MRS}, see also \cite{Larsen:1995ax,Rosenhaus:2014ula,Rosenhaus:2014zza}. In Appendix \ref{curvEE} we present an independent computation of the curvature correction to the universal `area law' and find full agreement with \reef{Sphi}. Note also that unlike odd $d$, the universal entanglement entropy in even dimensions is not affected by de Sitter temperature. Absence of thermal corrections is due to the fact that UV divergences of EE are state independent, see \eg \cite{Hung:2011ta}. 

So far we discussed how the logarithmic divergences of EE are encoded in \reef{dSdR}. In particular, we used a na\"{\i}ve bare expression \reef{Cphi} for $C_\phi$ to uncover the universal entanglement entropy. Counter terms which are unavoidable even in the absence of interactions will contribute to \reef{Cphi} and result in a finite $C_\phi$. 
This makes us believe that the universal entanglement entropy for a QFT in the setup under study is directly related to the logarithmic divergences associated with renormalization of the bare operators and bare parameters in the energy-momentum tensor trace, whereas renormalization of the theory generates a finite EE through the use of \reef{dSdR}. Of course, the finite part of EE depends on the choice of subtraction scheme, and therefore it is ambiguous, but the coefficients of the logarithmic running are universal. In what follows we elaborate the details of RG flow of the (finite) EE for a free massive scalar field in $d=2,4$. These results generalize easily to higher dimensions.

\subsubsection{Renormalization of EE in even d}
\label{sec:RenScalarEvend}
Eq. \reef{Sphi} reveals how the universal EE is encoded in \reef{dSdR} (recall that dimensional regularization eliminates non-universal power law divergences). However, as we stressed in the previous section, this calculation is incomplete without accounting for the counter terms which are necessary to render the partition function of the theory finite. Taking these terms into account results in a finite vev of the energy-momentum tensor and thus finite EE through the use of \reef{dSdR}. Of course, the universal divergence of EE is not lost. It transmutes into a finite term which represents the logarithmic running of EE. In this section we calculate all  necessary counter terms needed to renormalize the partition function and obtain a renormalized expression for the EE. 

\subsubsection*{ $d=2$ }

In $d=2$, the total action which includes all possible counter terms that are necessary to render the partition function of the theory finite is\footnote{For a free field theory the mass and the non-minimal coupling are not corrected by the divergences of higher-order loops, therefore we do not distinguish between the bare and renormalized $m^2$ and $\xi_c$.}
\be
S_\mt{tot}^{d=2}=\int_{S^d} \( {1\over 2}(\del\phi)^2 + {1\over 2} m^2\phi^2 + {1\over 2} \xi_c \cR \, \phi^2 + \Lambda_0 + \kappa_0 \cR \)~,
\ee
where $\Lambda_0$ describes a bare cosmological constant and $\kappa_0$ is the usual bare coupling of  the Einstein action. In 2D this term is responsible for the trace anomaly. The conformal coupling $\xi_c$ vanishes in two dimensions, however, we are going to use the method of dimensional renormalization, and therefore we keep it in the action.

The bare and renormalized parameters in the minimal subtraction scheme are related as 
\bea
 \Lambda_0&=&\mu^{d-2}\(\Lambda + {\Lambda_p\over d-2}\, m^2 \)~,
 \non
 \kappa_0&=&\mu^{d-2}\(\kappa + {\kappa_p\over d-2}\)~,
 \labell{bare2}
\eea
where $\Lambda,\kappa$ are renormalized parameters, $\mu$ is an arbitrary mass scale and $\Lambda_p$, $\kappa_p$ are residues of simple poles. These counter terms are necessary to ensure a finite partition function.  For an interacting QFT the above expression contains higher order poles and much more complicated residues.

The constant $\Lambda_p$ is determined from the requirement that 
\be
 - m^2 {\del Z \over \del m^2}=  \int_{S^d} \big( {m^2\over 2}\langle \phi^2\rangle + \mu^{d-2} m^2 {\Lambda_p\over d-2}\big)
\ee
must be finite. Expanding \reef{vacphi2} around $d=2$ gives
\be
 \langle \phi^2\rangle =-{1\over 2\pi(d-2)}+{1\over 4\pi}\( \log(4\pi R^2)-\psi(1-\lambda)-\psi(\lambda)-\gamma\) + \mathcal{O}(d-2)~.
 \labell{phi2}
\ee
Hence,
\be
 \Lambda_p={1\over 4\pi} ~.
 \labell{lamp}
\ee

Similarly, the free parameter $\kappa_p$ is determined by the divergence of a 1-loop diagram with external gravitons. However, we resort to a different method. Let us consider the vacuum expectation value of the energy-momentum tensor trace
\be
 \langle T(x) \rangle= - {2\over\sqrt{g}} g^{\mu\nu} {\delta\,\log Z \over \delta g^{\mu\nu}(x)} 
 =-{d-2\over 2} \langle\phi(-\nabla^2+\xi_c\cR+m^2)\phi\rangle-m^2 \langle\phi^2\rangle -d\Lambda_0-(d-2)\kappa_0\, \cR~.
\ee
The first term on the right hand side vanishes since it corresponds to the equation of motion operator's vev. The rest can be expanded around $d=2$ using \reef{bare2}, \reef{phi2}, \reef{lamp}
\be
  \langle T(x) \rangle=-{m^2\over 4\pi}\( \log(4\pi R^2\mu^2)-\psi(1-\lambda)-\psi(\lambda)-\gamma+1\) - 2\Lambda  -\kappa_p \, \cR~.
\ee

As expected, the vev of the energy-momentum tensor trace is finite after the bare operators and bare parameters are replaced with renormalized parameters. The unspecified $\mu, \Lambda$ and $\kappa_p$ can be determined by imposing the decoupling condition $\langle T\rangle\to 0$ as $mR\to\infty$ \cite{Forte:1998dx}
\be
 \mu^2=m^2\, , \quad \Lambda=m^2\,{(\gamma-\log(4\pi) -1) \over 8\pi} \, , \quad \kappa_p={-1\over 24\pi} ~.
\ee
Substituting into $\langle T\rangle$, we obtain 
\be
  \langle T(x) \rangle=-{m^2\over 4\pi}\( 2\log(mR)-\psi(1-\lambda)-\psi(\lambda)\)  + {\cR\over 24\pi}~.
\ee
Note that the last term cannot be modified by adding to the action a finite counter term, hence it represents the trace anomaly in two dimensions.

Next we evaluate the EE flow using \reef{1pT} and \reef{dSdR}
\bea
 C_\phi&=&{(mR)^2\over 4\pi}\( 2\log(mR)-\psi(1-\lambda)-\psi(\lambda)\)  - {1\over 12\pi}~,
 \non
 R{dS_\mt{EE}\over dR}&=&(mR)^2+{i (mR)^4\big(\psi'(1-\lambda)-\psi'(\lambda)\big)\over 2 \sqrt{(mR)^2-1/4} } + {1\over 3} ~.
\eea
It can be readily verified that $RS_\mt{EE}'$ vanishes as $mR\to \infty$. In this limit the RG running takes us to an empty fixed point. In the opposite limit, $mR\to 0$, we recover a pathological massless scalar field on $S^2$ ($\xi_c=0$ in 2D)\footnote{We call this theory `pathological' since Laplace operator on a sphere is not invertible. It has a normalizable zero mode which corresponds to a constant field configuration. Thus a minimally coupled massless scalar field on a sphere does not have a well-defined two point function without excluding zero mode from consideration.}. Therefore the above equations describe RG running of a massive theory into an empty fixed point.

\subsubsection*{ $d=4$ }

This case is similar to $d=2$ except that the total action admits more counter terms, see \eg \cite{Brown:1980qq},
\bea
 S_\mt{tot}^{d=4}&=&\int_{S^d} \( {1\over 2}(\del\phi)^2 + {1\over 2} m^2\phi^2 + {1\over 2} \xi_c \cR \, \phi^2 \right.
 \non
 &&\quad\quad\quad\quad\quad~\,
 +\left. \Lambda_0 + \kappa_0 \cR +{b_0\over 16\pi^2}C^{\mu\nu\rho\sigma}C_{\mu\nu\rho\sigma} +2a_0E_4+ c_0 \cR^2\)~,
\eea
where $C_{\mu\nu\rho\sigma}$ is the Weyl tensor. The relation between bare and renormalized parameters is
\bea
 \Lambda_0&=&\mu^{d-4}\(\Lambda + {\Lambda_p\over d-4}\, m^4 \)~,
 \quad
 \kappa_0=\mu^{d-4}\(\kappa + {\kappa_p\over d-4}m^2\)~,
  \labell{bare4}
  \\[1em]
 a_0&=&\mu^{d-4}\( a + {a_p\over d-4}\) ~, \quad
 b_0=\mu^{d-4}\( b + {b_p\over d-4}\)~, \quad 
 c_0=\mu^{d-4}\( c + {c_p\over d-4}\)~.
\nonumber
\eea

As before, the constant $\Lambda_p$ is obtained by requiring that 
\be
 - m^2 {\del Z \over \del m^2}=  \int_{S^d} \big( {m^2\over 2}\langle \phi^2\rangle + \mu^{d-4}m^2\cR{\kappa_p\over d-4} + 2 \mu^{d-4} m^4 {\Lambda_p\over d-4} \big)
\ee
is finite. Expanding \reef{vacphi2} around $d=4$ gives
\be
 \langle \phi^2\rangle ={m^2\over 8\pi^2(d-4)}-{m^2\over 16\pi^2}\big( \log(4\pi R^2)-\psi(\lambda)-\psi(3-\lambda)+1-\gamma\big) + \mathcal{O}(d-4)~.
 \labell{4Dphi2}
\ee
Hence,
\be
 \Lambda_p={-1\over 32\pi^2} ~, \quad \kappa_p=0~.
 \labell{lamp4}
\ee

To determine other counter terms, let us consider vev of the energy-momentum trace,
\bea
 \langle T(x) \rangle
 &=&-{d-2\over 2} \langle\phi(-\nabla^2+\xi_c\cR+m^2)\phi\rangle-m^2 \langle\phi^2\rangle -d\Lambda_0-(d-2)\kappa_0\, \cR
 \non
 &-&(d-4)\( {b_0\over 16\pi^2}C^{\mu\nu\rho\sigma}C_{\mu\nu\rho\sigma} +2 a_0E_4+ c_0 \cR^2 \)+4(d-1)c_0 \nabla^2\cR~.
\eea
The first term is proportional to the equation of motion operator's vev, thus it vanishes. Furthermore, $\cR$ is constant on a sphere, whereas Weyl tensor vanishes. Thus,
\bea
 \langle T(x) \rangle
 &=&-m^2 \langle\phi^2\rangle -d\Lambda_0-(d-2)\kappa_0\, \cR-(d-4)\( -2 a_0 E_4+ c_0 \cR^2 \)
 \\
 &=&{m^4\over 16\pi^2}\big( \log(4\pi R^2\mu^2)-\psi(\lambda)-\psi(3-\lambda)+{3\over 2}-\gamma\big)-4\Lambda-2\kappa\cR
 -2a_pE_4 - c_p\cR^2 ~,
 \nonumber
\eea
where we used \reef{bare4}, \reef{4Dphi2} and \reef{lamp4}. The last two terms correspond to the trace anomaly. Obviously, these terms cannot be modified by adding a finite counter term to the action. 

Next we impose the decoupling condition, $\langle T\rangle\to 0$ as $mR\to\infty$, to determine $a_p, c_p, \mu, \kappa, \Lambda$ \cite{Forte:1998dx}
\be
 \mu^2=m^2\, , \quad \Lambda=m^4\,{3-2\gamma+2\log(4\pi)  \over 128\pi^2} \, , \quad \kappa={-m^2\over (24\pi)^2} ~, \quad {3\over 2\pi^2}a_p+144\, c_p={1\over 240\pi^2}~.
\ee

Although for our needs it is not necessary to calculate $a_p$ and $c_p$ separately, it is still worth mentioning that $c_p=0$ for a free massive scalar field. Indeed, $c_p$ is mass independent, and it vanishes for a massless conformally coupled scalar field (CFT). One can also verify that $c_p=0$ by a direct calculation \cite{Brown:1976wc,Brown:1980qq}. Substituting the above parameters into $\langle T\rangle$, we obtain
\be
 \langle T(x) \rangle ={m^4\over 16\pi^2}\( 2\log(m R)-\psi(\lambda)-\psi(3-\lambda) + {2\over 3(mR)^2}\) - {1\over 240\pi^2 R^4} ~.
\ee
Finally, using \reef{1pT} and \reef{dSdR}, results in
\bea
 C_\phi&=&-{(mR)^4\over 16\pi^2}\( 2 \log(m R)-\psi(\lambda)-\psi(3-\lambda) + {2\over 3(mR)^2}\) + {1\over 240\pi^2}~,
 \non
 R{dS_\mt{EE}\over dR}&=&{(mR)^2\over 18}-{(mR)^4\over 12} -{i (mR)^6\big(\psi'(3-\lambda)-\psi'(\lambda)\big)\over 24 \sqrt{(mR)^2-1/4} } - {1\over 90} ~.
\eea

In the deep IR limit, $mR\to \infty$, the theory flows into an empty fixed point and $RS_\mt{EE}'$ vanishes. In the UV limit, $mR\to 0$, we recover a conformally coupled massless scalar field, and $RS_\mt{EE}'=-1/90$ in accordance with \reef{EEanom}.\footnote{$a=a_p={1\over 360}$.} Therefore RG running happens between two fixed points which correspond to the conformally coupled scalar field and an empty theory.

\subsection{Dirac fermion}

Let us consider a free Dirac field of mass $m$. The Euclidean action is given by
\be
I=\int d^dx \, \bar\psi\( \slashed\nabla+m\)\psi~,
\ee
and the corresponding energy-momentum tensor reads
\be
T_{\mu\nu}={1\over 2} \bar\psi\gamma_{(\al}\overset{\leftrightarrow}{\nabla}_{\bt)}\psi-\delta_{\al\bt}(\bar\psi\slashed\nabla\psi+m\bar\psi\psi) ~,
\ee
where $\gamma_\mu$ are the gamma matrices satisfying the anticommutation relations
\be
\{\gamma_\mu,\gamma_\nu\}=2g_{\mu\nu}~.
\ee  
Taking the trace of the energy-momentum tensor and using the Dirac equation of motion we obtain,
\be
T=-m\,\bar\psi\psi.
\ee 
Hence
\be
C_\psi=m\,R^d \langle\bar\psi\psi\rangle ~.
\ee
To evaluate $C_\psi$ we use propagator of the Dirac field on $S^d$ \cite{Camporesi:1992tm}
\bea
\langle \psi(y) \bar\psi(x)\rangle&=& {\Gamma\big({d\over 2}+i mR \big) \Gamma\big({d\over 2} - i mR\big)\over 2^{d}\pi^{d/2}\Gamma\big({d\over 2}+1\big) R^{d-1}}
\\
&\times&\(mR \, U(y,x) \cos{\chi\over 2} 
\ _{2}F_1\Big({d\over 2}+i mR~,~ {d\over 2} - i mR~;~{d\over 2}+1~;~\cos^2{\chi\over 2}\Big) 
\right.
\non
&+&\left.{d\over 2} n^\mu\gamma_\mu U(y,x) \sin{\theta\over 2}  \ _{2}F_1\Big({d\over 2}+i mR~,~ {d\over 2} - i mR~;~{d\over 2}~;~\cos^2{\chi\over 2}\Big)   \)~,
\nonumber
\eea
where $n^\mu$ is the unit tangent vector to the geodesic connecting $x$ to $y$, $U(y,x)$ is a matrix in the spinor indices which parallel propagates a spinor between the two points, and $\chi$ is the polar angle between $x$ and $y$. In particular,
\be
\langle\bar\psi\psi\rangle= 
-2^{\[{d\over 2}\]}{\Gamma\big({d\over 2}+i mR \big) \Gamma\big({d\over 2} - i mR\big) \sinh(\pi m) \over R^{d-1} (4\pi)^{d/2} \sin({d\pi\over 2}) \Gamma\({d\over 2}\)}~.
\ee
This vev is finite for odd $d$, whereas for even $d$ it has simple poles which correspond to the logarithmic divergences. Let us consider these cases separately

\subsubsection*{Odd $d$}

For odd $d$, we have
\bea
C_\psi&=&(-)^{d+1 \over 2} {(\pi mR) \tanh(\pi mR) \over \sqrt{2}\,(2\pi)^{d/2} \Gamma\({d\over 2}\) } \prod_{j=1}^{d-1\over 2} \( (d/2-j)^2+m^2R^2\)
\\
&=&(-)^{d+1 \over 2} {(\pi mR) \tanh(\pi mR) \over \sqrt{2}\,(2\pi)^{d/2} \Gamma\({d\over 2}\) } 
\((mR)^{d-1} + {d(d-1)(d-2)\over 24} (mR)^{d-3} + \ldots +{\Gamma^2\big({d\over 2}\big)\over \pi} \)
\nonumber
\eea
The hyperbolic function is associated with the thermal corrections, and in principle it should be stripped off to isolate the impact of entanglement. We achieve this goal by taking the limit $mR\gg 1$, in which case the thermal effect is exponentially suppressed. To leading order we recover a known result \cite{Hertzberg:2010uv}, see also Appendix \ref{freeflat}
\begin{flalign}
C_\psi &\underset{\mt{$mR\gg 1$}}{=}
{\pi  (-)^{d+1 \over 2}  \over \sqrt{2}\,(2\pi)^{d/2} \Gamma\({d\over 2}\) } 
\((mR)^d + {d(d-1)(d-2)\over 24} (mR)^{d-2} + \ldots +{\Gamma^2\big({d\over 2}\big)\over \pi} mR\)\\
R{d S_\mt{univ}^\mt{Dirac}\over dR} 
&\underset{\mt{$mR\gg 1$}}{=} { (-)^{d-1 \over 2}  \pi (d-2) \over 12 \sqrt{2}\,(2\pi)^{d-2\over 2} \Gamma\({d\over 2}\) }  \, A_\Sigma \, m^{d-2} 
+ \ldots ~,
\end{flalign}
where ellipsis encapsulate curvature corrections.

\subsubsection*{Even d}
The pole structure of $C_\psi$ for even $d_0$ is given by
\begin{flalign}
C_\psi&=(-)^{d_0/2-1}\,{\prod_{j=1}^{d_0/2} \( (d_0/2-j)^2+m^2R^2\)  \over (2\pi)^{d_0/2} \Gamma\({d_0\over 2}\)} {2\over d-d_0} + \ldots
\non
&={ (-)^{d_0/2-1} \over (2\pi)^{d_0/2} \Gamma\({d_0\over 2}\)} \((mR)^{d_0} + {d_0(d_0-1)(d_0-2)\over 24} (mR)^{d_0-2} \right.
\\
&+\left.\frac{d_0 (d_0-1) (d_0-2) (d_0-3) (d_0-4) (5d_0+2)}{5760}(mR)^{d_0-4} +\ldots + \Gamma^2\big({d_0\over 2}\big) (m R)^2  \){2\over d-d_0}+\ldots 
\nonumber
\end{flalign}
The universal entanglement entropy is obtained by substituting it into \reef{dSdR}
\bea
R {d S_\mt{univ}^\mt{Dirac} \over dR}&=& { (-)^{d_0/2} (d_0-2) A_{\Sigma} \over 6 (2\pi)^{(d_0-2)/2} \Gamma\({d_0\over 2}\)} 
\(  m^{d_0-2} + \frac{(d_0-3) (d_0-4) (5d_0+2)}{120}\frac{m^{d_0-4}}{R^2} \right.
\non
&+&\left. \ldots + 12 {\Gamma^2\big({d_0\over 2}\big) \over d_0(d_0-1)}{m^2\over R^{d-4}}\)\log\delta
\labell{Spsi}
\eea
The leading order term matches a well-known universal `area law' \cite{Hertzberg:2010uv}, see also Appendix \ref{freeflat}. Subleading terms represent corrections to the entanglement entropy induced by curvatures. In Appendix \ref{curvEE} we present independent computation of the curvature corrections and find full agreement with \reef{Spsi}. Although we do not present it here, similar analysis to section \ref{sec:RenScalarEvend} for the massive scalar can be applied to the massive fermion, giving a renormalized expression for the EE, and the UV limit agrees with \reef{EEanom}.

\section{Spectral decomposition}
\labell{sec:spec}

In this section we derive an expression for the right hand side of \reef{flow2} in terms of specific spectral function. We start from the spectral decomposition of the conserved two point function \reef{conTT}. In general, it can be expressed as \cite{Osborn:1999az}
\be
 \Omega_{d-1}^2 \langle T_{\mu\nu}(x) T_{\al\bt}(y) \rangle_\mt{con}=\Gamma_{0\mu\nu,\al\bt}(x,y)+\Gamma_{2\mu\nu,\al\bt}(x,y)~,
 \labell{specrep}
\ee
where the spin-2 piece, $\Gamma_{2\mu\nu,\al\bt}$, is traceless, and therefore it does not contribute to \reef{flow2}.\footnote{Explicit expression for $\Gamma_{2\mu\nu,\al\bt}$ is derived in \cite{Osborn:1999az}.}. The spin-0 piece is given by
\bea
 \Gamma_{0\mu\nu,\al\bt}(x,y)&=&S_{\mu\nu}(x) F_0(\sigma) \overleftarrow{S}_{\al\bt}(y)~,
 \non
 S_{\al\bt}&\equiv&\nabla_\al\nabla_\bt- g_{\al\bt}\nabla^2-{(d-1)\over R^2} g_{\al\bt}~,
\eea
where $\sigma(x,y)$ is the geodesic interval in units of $R$ between $x$ and $y$, covariant derivatives on the left of $F_0(\sigma)$ act on $x$, whereas covariant derivatives on the right of $F_0(\sigma)$ act on $y$, as indicated by the arrow sign above $S_{\al\bt}(y)$, and
\be
 F_0(\sigma)=\Omega_{d-1} \int_{\mu_\phi}^\infty d\mu\rho_0(\mu)G_{\mu}(\sigma)~,
 \labell{F0}
\ee
where as before $G_\mu(\sigma)$ is the Green's function satisfying 
\be
 \big(-\nabla^2+\xi_c\cR+\mu^2 \big) G_\mu(x,y)=\delta^d(x,y)~,
 \label{green}
\ee
with $\mu$ being the mass of the field. Thus, the two point function which is necessary to evaluate the right hand side of \reef{flow2} is completely determined by $\rho_0(\mu)$,
\be
 \Omega_{d-1}^2 \langle T(x) T_{\al\bt}(y) \rangle_\mt{con}=g^{\mu\nu}S_{\mu\nu}(x) F_0(\sigma) \overleftarrow{S}_{\al\bt}(y)~.
 \labell{TT}
\ee

Before substituting \reef{TT} into \reef{flow2}, it is worth mentioning that for our choice of foliation \reef{coor}, we have $\xi^{\mu}=\delta^\mu_\tau$ and $n_\mu=R\cos\theta\,\delta^\tau_\mu$. Hence,
\bea
\nabla^2&=&g^{\tau\tau}\nabla_\tau\nabla_\tau+{1\over R^2\sin^{d-2}\theta}\del_\theta\big( \sin^{d-2}\theta\del_\theta\big)+ {1\over R^2\sin^2\theta}\nabla^{2}_{S^{d-2}}
\non
 \xi^\al n^\bt S_{\al\bt}&=& 
 -R\cos\theta \({1\over R^2\sin^{d-2}\theta}\del_\theta\big( \sin^{d-2}\theta\del_\theta\big)+ {1\over R^2\sin^2\theta}\nabla^{2}_{S^{d-2}}+{(d-1)\over R^2} \)~,
 \labell{xinS}
\eea
with $\nabla^{2}_{S^{d-2}}$ being the intrinsic Laplacian on $S^{d-2}$.

Next, we find it useful to define
 \be
  G_{\mu}^{(2)}(\tau-\tau',\theta,\theta')\equiv \int d\Omega' ~ G_{\mu}(\tau,\theta,\Omega;\tau',\theta',\Omega')
  \labell{G2}
 \ee
where $(\tau,\theta,\Omega)$ and $(\tau',\theta',\Omega')$ are coordinates \reef{foliation} of $x$ and $y$ respectively, and the integral runs over a $(d-2)$-dimensional sphere parametrized by $\Omega'$. The first thing to note about $G_{\mu}^{(2)}$ is that it depends on the difference $\tau-\tau'$ due to the symmetry of \reef{foliation} under translations in $\tau$, and it is independent of $\Omega$ since \reef{foliation} is symmetric under rotations of $S^{d-2}$. 

Combining \reef{entham} and \reef{TT}, leads to
\bea
  \langle T(x) K \rangle_\mt{con}&=& -{2\pi(d-1)\over \Omega_{d-1}} \int_V \int_{\mu_\phi}^\infty d\mu \big(\xi_c\cR + {d\over R^2} + \mu^2 \big) \rho_0(\mu)G_{\mu}(x,y)
 \overleftarrow{S}_{\al\bt}(y)\xi^\al n^\bt 
 \non
 &=& {2\pi(d-1)\over \Omega_{d-1}} \int_{\mu_\phi}^\infty d\mu \big(\xi_c\cR + {d\over R^2} + \mu^2 \big) \rho_0(\mu) 
    \labell{TK} \\
 &&\times
 \int_0^{\pi\over 2}d\theta' \cos\theta'\(
 \del_{\theta'}\big( \sin^{d-2}\theta' \del_{\theta'}\big)+(d-1)\sin^{d-2}\theta' \) G_{\mu}^{(2)}(\tau,\theta,\theta')~,
\nonumber
\eea
where based on \reef{green} we substituted $\nabla^2G_{\mu}(\sigma)=(\xi_c\cR+\mu^2)G_{\mu}(\sigma)$ in the first equality,\footnote{We ignored the delta function on the right hand side of \reef{green}. There is nothing bad about it if $x$ is disjoint from $V$. If, however, $x$ hits the support of $K$, then one may question whether it is justified to suppress the delta function in \reef{green}. In general, if supports of the operators overlap, it is a must to include contact terms. However, inclusion of such contact terms will break the $O(2)$ invariance inherent to the setup under study, and therefore  we disregard them. From computational point of view, we simply assume that $x\notin V$.} whereas in the second equality we used \reef{xinS} and the definition \reef{G2}.\footnote{Note that by assumption $x$ is disjoint from $V$, therefore $G_\mu(x,y)\overleftarrow{\nabla}^2_{S^{d-2}}$ is everywhere regular on $V$ and its integral over $S^{d-2}$ vanishes just because this manifold has no boundaries.} We also assumed that $d>2$.\footnote{The special case $d=2$ can be treated similarly. In this case the entangling surface consists of two disjoint points $\theta'=\pm{\pi\over 2}$, and the integral over $\theta'$ in \reef{TK} should be extended from $-{\pi\over 2}$ to ${\pi\over 2}$ in order to cover all of $S^2$, see \reef{coor}.}  Finally, integrating the right hand side of \reef{TK} by parts twice, yields
\bea
 \langle T(x) K \rangle_\mt{con}&=& {2\pi(d-1)\over \Omega_{d-1}} \int_{\mu_\phi}^\infty d\mu \big(\xi_c\cR + {d\over R^2} + \mu^2 \big) \rho_0(\mu) 
 G_{\mu}^{(2)}(\tau,\theta,\theta'={\pi\over 2})~,
    \labell{TK2}
\eea
where $G_{\mu}^{(2)}(\tau,\theta,\theta'={\pi\over 2})$ is independent of $\tau$ since $\theta'={\pi\over 2}$, and the system exhibits rotational symmetry in the transverse space. As expected, $\langle T(x) K \rangle$ is just a function of $\theta$ which parametrizes the geodesic distance from the entangling surface in the transverse space.

Another useful result follows directly from \reef{green}
\be
  \int d^dx \, \sqrt{g(x)} G_\mu(x,y)={1\over \xi_c\cR+\mu^2 }~.
  \labell{intGreen}
\ee
Indeed, the integral on the left hand side is convergent since the Green's function is everywhere regular on a sphere except at $x=y$, where it diverges as $\sigma^{2-d}$. However, this divergence is balanced by the integration measure which behaves as $\sigma^{d-1}$. Furthermore, $S^d$ is a maximally symmetric space, and therefore the result of integration is given by some constant independent of $y$. In particular, \reef{intGreen} follows from the following identities
\be
  (\xi_c\cR+\mu^2) \int d^dx \, \sqrt{g(x)} G_\mu(x,y)=(-\nabla^2+\xi_c\cR+\mu^2) \int d^dx \, \sqrt{g(x)} G_\mu(x,y)=1~,
\ee
where the last equality rests on \reef{green}.

Substituting now \reef{TK2} into \reef{flow2}, and using \reef{G2}, \reef{intGreen} to integrate $G^{(2)}_\mu$over $x$, yields
\be
 R {dS_\mt{EE} \over dR} =  -  {(d-1) \Gamma\big({d\over2}\big)\over \pi^{d-2\over 2}} \mathcal{A}_\Sigma\int_{\mu_\phi}^\infty d\mu \(1 + {d\over R^2(\xi_c\cR+ \mu^2)} \) \rho_0(\mu)
 ~,
 \labell{Sprime}
\ee
where $\mathcal{A}_\Sigma$ is the area of the entangling surface. 

The integral on the right hand side of \reef{Sprime} is convergent at the lower bound since sphere introduces a natural IR cut off, but it does not necessarily converges at the upper bound. Indeed, in the UV limit curvature corrections are irrelevant, and convergence of the integral is essentially the same as in Minkowski space. Note also that in the limit $\mu^2\cR\gg1$ we recover the result of \cite{Casini:2014yca}.

Furthermore, if the integral on the right hand side converges, then positivity of the spectral function guarantees $R S_\mt{EE}' \leq 0$ along the RG flow. However, it certainly does not ensure \reef{Fthm} since finite EE at the IR fixed point, $S_\mt{EE}^\mt{IR}$, includes terms which are part of the UV physics. These terms should be subtracted to isolate the universal contribution, $c_0^\mt{IR}$. We elaborate details of this point in the discussion section.

\subsection{Example: conformally coupled scalar on $S^3$}

In this section we carry out spectral decomposition for a massive conformally coupled scalar field described by the Euclidean action \reef{freeS}. In this case, based on \reef{imptr}, we have 
\be
 \langle T(x) T(y) \rangle = m^4 \langle \phi^2(x) \phi^2(y) \rangle = 2 m^4 G^2_m(x,y)~,
 \labell{trace2p}
\ee
where by assumption $x$ and $y$ are two disjoint points, and therefore the equation of motion operator in \reef{imptr} does not contribute. In particular, it follows that the spin-0 piece of  \reef{specrep} is straightforwardly related to the spectral representation of 
\be
 \langle \phi^2(x) \phi^2(y) \rangle =  \int_{\mu_\phi}^\infty d\mu \, \rho_{\phi^2}(\mu) G_\mu(x,y)
 =\int_{\mu_\phi^2}^\infty d\mu^2 \, \tilde\rho_{\phi^2}(\mu) G_\mu(x,y)~,
 \labell{spectr}
\ee
where $\rho_{\phi^2}(\mu)=2\mu\tilde\rho_{\phi^2}(\mu)$. In the meantime we keep the lower bound $\mu_\phi$ unspecified. We find it convenient to use $\mu^2$ as the integration measure in the spectral representation since $\mu_\phi$ may admit imaginary values to account for the possibility of negative $\mu_\phi^2$ on a sphere.

A useful relation can be established between $\del_{m^2}\langle\phi^2\rangle$ and $\tilde\rho_{\phi^2}$ 
\be
 -2{\del\over\del m^2}\langle \phi^2 \rangle =\int d^d y\sqrt{g(y)} \langle \phi^2(y) \phi^2(x) \rangle = \int_{\mu_\phi^2}^\infty d\mu^2 
 { \tilde\rho_{\phi^2}(\mu) \over \xi_c\cR+ \mu^2 }~,
\ee
where in the last equality we used \reef{intGreen}. Shifting the integration measure, $z=(R\mu)^2-(R\mu_\phi)^2$ and using \reef{vacphi2}, yields
\be
  { \psi(\lambda)-\psi(d-1-\lambda)-\pi \cot\big({\pi\over 2}(d-2\lambda)\big) \over \cN}\,\langle \phi^2 \rangle 
  = i R^{\,2-d} \int_0^\infty dz 
 { \tilde\rho(z) \over (d-1)^2/4 + \cN_\phi^2+z }~.
 \labell{delphi2}
\ee
where $\cN_{\phi}^2=(R\mu_\phi)^2-{1\over 4}$ and $\tilde\rho_{\phi^2}=R^{\,4-d}\,\tilde\rho $. We will use this result to check our computations in what follows.

For the rest of this section we continue to explore the spectral decomposition in $d=3$. In this case  solution \reef{greensol} to the Green's equation can be expressed in terms of elementary functions
 \be
 G_m(\chi)=-{\sinh\[\cN(\chi-\pi)\] \over 4\pi R \, \sinh (\pi\cN )\, \sin\chi} ~, \quad\quad \cN\equiv \sqrt{ (mR)^2 - {1\over 4}}
 \quad ,
 \label{eqn:prop-on-sphere}
 \ee
We choose to work in the basis of scalar spherical harmonics on $S^3$ of radius $R$. They are given by \cite{Higuchi:1986wu}
 \be
 Y_{l_3,l_2,l_1}(\chi,\theta,\phi)={1 \over
 R^{3/2}\sqrt{2\pi}}e^{il_1\phi}\ _2 c_{l_2}^{l_1}\,P_{l_2}^{-l_1}(\cos\theta)\,
 _3 c_{l_3}^{l_2}\,(\sin\chi)^{-1/2}\,P_{l_3+{1\over 2}}^{-(l_2+{1\over2})}(\cos\chi)
 \quad ,
 \label{eqn:spherical-harmonics}
 \ee
where $l_3\geq l_2\geq |l_1|$ and
 \bea
 P^{-\mu}_{\nu}(x)&=&{1 \over \Gamma(1+\mu)}\( {1-x \over 1+x} \)^{\mu/2}
 \ _{2}F_1\(-\nu ~,~ \nu+1 ~;~1+\mu ~;~ {1-x\over 2}\)
 \quad ,
  \nonumber \\
 _n c_{L}^{l}&=&\[ {2L+n-1 \over 2}{(L+l+n-2)! \over (L-l)!} \]^{1/2}
 \quad .
 \label{eqn:associated-Legandre}
 \eea
Note, that by definition spherical harmonics satisfy
\be
 Y_{l_3,l_2,l_1}^*(\chi,\theta,\phi)=(-1)^{l_1}Y_{l_3,l_2,-l_1}(\chi,\theta,\phi)~.
 \labell{Ystar}
\ee 
In particular, the only nontrivial coefficients in the expansion of $G_m(\chi)$, are given by
\be
 G_{l_3}(m^2)=\int_{S^3} Y_{l_300}(\chi)G_m(\chi)={R^{1/2}\over \sqrt{2}\pi} {l_3+1\over (l_3+1)^2 + \cN^2}~,
 \labell{GL}
\ee
where according to \reef{eqn:spherical-harmonics}
 \be
 Y_{l_300}(\chi)={1 \over R^{3/2}\sqrt{2}\pi}{\sin\[(l_3+1)\chi\]\over\sin\chi}
 \quad .
 \label{harmon2}
 \ee
Similarly, one can find the expansion of $G^2_m(\chi)$. 

According to (\ref{eqn:prop-on-sphere}),
 \be
 G^2_m(\chi)={\sinh^2\[\cN(\chi-\pi)\] \over 16\pi^2 R^2 \sinh^2\(\pi\cN\)\sin^2\chi}
 \quad .
 \ee
Hence, its expansion in terms of spherical harmonics (\ref{eqn:spherical-harmonics}), is given by
\begin{multline}
 G_{l_3}^{(2)}(m^2)=\int_{S^3} Y_{l_300}(\chi)G^2_m(\chi)=
 {1 \over 2^{5/2}\pi^2\,R^{1/2} \sinh^2\(\pi\,\mathcal{N}\)}
 \\ \times
 \left\{\begin{array}{ll}
   2\mathcal{N} \sinh(2\pi \mathcal{N})\( {1 \over 1+4\mathcal{N}^2}+{1 \over 9+4\mathcal{N}^2}+\cdot\cdot\cdot
   +{1 \over l_3^2+4\mathcal{N}^2}\)\quad\quad\quad\,;l_3 \quad \mathrm{odd} \\
   -{\pi \over 2}+{\sinh(2\pi \mathcal{N}) \over 4}\( {1 \over \mathcal{N}}+ {2\mathcal{N} \over 1+\mathcal{N}^2}
   + {2\mathcal{N} \over 4+\mathcal{N}^2}+\cdot\cdot\cdot+{2\mathcal{N} \over (l_3/2)^2+\mathcal{N}^2}\)
   ;l_3 \quad \mathrm{even}
 \end{array}\right.
 \label{Acoeff}
\end{multline}
Using the digamma function, $\psi(x)$, it can be succinctly written as
\be
G_{l_3}^{(2)}(m^2)={1\over 2^{7/2} R^{1/2} \pi}\[  1-i\,{\coth(\pi\cN)\over\pi}\, \big[\psi\big(1+{l_3\over 2} -i\cN\big)-\psi\big(1+{l_3\over 2} +i\cN\big) \big] \]~.
\label{G2L}
\ee

The spectral representation \reef{spectr} is equivalent to the spectral representation in the angular momentum space
\be
 G_{l_3}^{(2)}(m^2)= {1\over 2}\int_{\mu_\phi^2}^\infty d\mu^2 \, \tilde\rho_{\phi^2}(\mu) G_{l_3}(\mu^2)~.
\ee
Substituting \reef{GL} and \reef{G2L}, and shifting the integration variable $z=(R\mu)^2-(R\mu_\phi)^2$, yields
\begin{multline}
{1\over 4 (l_3+1)} \[  1-i\,{\coth(\pi\cN)\over\pi}\, \big[\psi\big(1+{l_3\over 2} -i\cN\big)-\psi\big(1+{l_3\over 2} +i\cN\big) \big] \]
\\
=\int_{0}^\infty dz \,  {\tilde\rho(z)\over \big(l_3+1\big)^2 + \cN_{\phi}^2 +z}~.
\labell{spectreq}
\end{multline} 
As expected, the special case of this formula, $l_3=0$, matches \reef{delphi2}.

Before we proceed, let us discuss possible values of $\mu_\phi^2$. If $\mu_\phi^2=0$ it means that the spectrum of particles starts from a conformally coupled scalar field ($z=0$ in this case) and continues all the way up to infinite massive modes ($z=\infty$). However, there is nothing bad about negative $\mu_\phi^2$ since conformal coupling in the Euclidean action \reef{freeS} may compensate its negativity such that the overall $\phi^2$ term is positive. The lowest possible negative value of $\mu_\phi^2$ is given by $(R\mu_\phi)^2=-{3\over 4}$. It can be read off either from the Euclidean action \reef{freeS}, or by setting $z=l_3=0$ in the integrand of \reef{spectreq} and demanding positivity. 

We need to invert \reef{spectreq} to get the spectral density $\tilde\rho(z)$. For brevity, we define
\bea
 z_0&\equiv&\big(l_3+1\big)^2 + \cN_{\phi}^2  \quad \Rightarrow \quad l_3= -1 +\sqrt{z_0-\cN_{\phi}^2} \,,
\non
f(z_0)&\equiv&\int_{0}^\infty dz \,  {\tilde\rho(z)\over z_0 +z}~,
\labell{not}
\eea
To express $\tilde\rho(z)$ in terms of $f(z)$, we first note that by definition
\be
\lim_{\epsilon\to 0}{ f(-z_0-i\epsilon) - f(-z_0+i\epsilon) \over 2i}= \lim_{\epsilon\to 0} \int_{0}^\infty dz \,  \tilde\rho(z)  {\epsilon \over (z-z_0)^2 +\epsilon^2}
\ee
Combining this result with 
\be
 \delta(z-z_0)={1\over\pi}\lim_{\epsilon\to 0}{\epsilon \over (z-z_0)^2 +\epsilon^2}~,
\ee
yields,
\be
\tilde\rho(z)={1\over \pi} \lim_{\epsilon\to 0}{ f(-z-i\epsilon) - f(-z+i\epsilon) \over 2i}~.
\ee

Using now \reef{spectreq}, \reef{not} and the definition of $\tilde\rho$,  we get,
\begin{multline}
 \rho_{\phi^2}(\mu)= {\mu R \over 2 \pi \cN_\mu} \left(  1-i\,{\coth(\pi\cN)\over2\pi}\, \left[\psi\big({1\over 2}(1-i\cN_\mu) -i\cN\big) -\psi\big({1\over 2}(1-i\cN_\mu) +i\cN\big) \right. \right.
 \\
 \left. \left.+\psi\big({1\over 2}(1+i\cN_\mu) -i\cN\big)-\psi\big({1\over 2}(1+i\cN_\mu) +i\cN\big)\right] \right)\Theta\big( (\mu R)^2 - 1/4 \big)
 \labell{specfunc}
\end{multline}
where $\cN_\mu=\sqrt{(\mu R)^2 - 1/4 }$. There are certain limits when \reef{specfunc} can be checked. 
\begin{itemize}
\item{
In the limit of conformally coupled scalar which is also a UV limit, $mR \to 0$, and we have $\cN={i\over 2}$. Hence, in this limit
\be
 \rho_{\phi^2}(\mu) \underset{mR\to0}{\to} {\mu R \over 2 \pi \sqrt{(\mu R)^2-1/4 }} \Theta\big( (\mu R)^2 - 1/4 \big)~.
\ee
and it can be checked by a direct computation that \reef{spectreq} holds. 
}
\item{
Flat space limit is recovered if we fix $m$ and $\mu$ while $R\to\infty$, \ie $R\gg m^{-1}, \mu^{-1}$
\be
 \rho_{\phi^2}(\mu) \underset{R\to\infty}{\to} {1\over 2\pi} \Theta\big( \mu - 2m \big)~,
\ee
where we used $\psi(x)\sim\log(x)$ for $x\gg 1$. In particular, in accord with \cite{Cappelli:1990yc}, we obtain
\be
 \rho_0(\mu)  \underset{R\to\infty}{\to} {1\over 2} \({m\over \mu}\)^4\Theta\big( \mu - 2m \big)~.
\ee
Substituting into \reef{Sprime} gives $RS_\mt{EE}' \sim mR\gg 1$. On the other hand, according to \reef{flow2} one expects $RS_\mt{EE}' =0$ since $mR\to\infty$ limit represents deep IR in our setup, which corresponds to an empty theory in this case. This contradiction is, however, apparent. Indeed, $R$  represents the characteristic RG scale and therefore one should only consider $\mu\sim{1/R}\ll m$. In this region theta-function and hence $\rho_0$ vanish. On the other hand, the spectral integral \reef{Sprime} accounts for all possible scales including $\mu\gg m$, and therefore $RS_\mt{EE}'$ is contaminated by various terms which are not part of the IR physics.
}
\item{
It worth noting that in the limit of $\mu R\gg 1$ with $m$ and $R$ fixed, we expect that the spectral function on a sphere asymptotically approaches its counterpart in flat space. Indeed,
\be
 \rho_{\phi^2}(\mu) \underset{\mu R\gg 1}{\to} {1 \over 2 \pi }~.
\ee
}

\end{itemize}

\section{Discussion}

In this paper we study the renormalization group flow of entanglement entropy in an analytically tractable setup - a cap-like entangling region in de Sitter space. Our system has much in common with the indispensable Rindler wedge, \eg the entangling surface exhibits $O(2)$ symmetry in the transverse space, and the corresponding entanglement entropy equals thermal entropy for {\it any} QFT.  However, in contrast to the Rindler space, where the geometry is flat and there is no characteristic temperature, in our setup the background curvature does not vanish, and it sets a characteristic length scale, $R$, for both the curvature of the entangling surface and temperature of the environment. 

Since there is only one global geometric scale, it also effectively determines the mass scale of RG running to be of order $1/R$. In particular, studies of RG flow boil down to a constant Weyl rescaling of the background geometry. From this perspective, our setup is scalable \cite{Liu:2012eea}, and we argue that EE satisfies the following RG equation
\begin{equation}
R\frac{d S_{\mt{EE}}}{d R}  = - { V_{S^d} \over d} \,R{d\over dR} \langle T^\mu_\mu\rangle ~,
\end{equation}
with $V_{S^d}$ being the volume of a $d$-dimensional sphere. 

The above simple relation between the entanglement entropy flow and trace of the energy-momentum tensor allows to analytically explore various properties of the entanglement entropy running when the system flows between the UV and IR fixed points. 

In section \ref{free} we scrutinize the RG flows of EE in $d=2,3,4$ for a conformally coupled scalar field deformed by a mass operator. In $d=3$ our findings indicate that the renormalized entanglement entropy defined in \cite{Liu:2012eea,Casini:2012ei,Liu:2013una} exhibits stationarity at the fixed points on a sphere, but it is not monotonic along the RG trajectory. This result is completely opposite to the behaviour of REE in Minkowski space, where it has been shown that REE is monotonic \cite{Casini:2012ei}, but not stationary \cite{Klebanov:2012va}. 

Lack of monotonicity may result from various causes. For instance, REE in de Sitter space receives thermal corrections which are absent in Minkowski space. Furthermore, absence of IR divergences in a QFT living on a sphere might be another plausible explanation for a qualitative difference in the behaviour of REE on de Sitter and on Minkowski space. That being said these arguments are unlikely to explain the discrepancies in the behaviour of REE in the vicimity of UV fixed point, since in the deep UV both the curvature and temperature of de Sitter space are irrelevant. 

Moreover, our calculations reveal another interesting and closely related result. We find that the universal `area law' for a conformally coupled scalar field is different from the known expression in Minkowski space \cite{Hertzberg:2010uv,Huerta:2011qi,Lewkowycz:2012qr}. This discrepancy cannot be attributed to a curved geometry since the universal `area law' depends solely on the area of entangling surface. This observation raises a controversial \cite{Solodukhin:1996jt,Hotta:1996cq,Solodukhin:2011gn,Nishioka:2014kpa,Lee:2014zaa,Herzog:2014fra,Dowker:2014zwa,Casini:2014yca} question whether there exists any difference  between the universal entanglement entropies for minimally and conformally coupled scalar fields in {\it flat} space. Possible explanation and interpretation of this phenomenon can be found in \cite{MRS} and  \cite{Larsen:1995ax,Rosenhaus:2014ula,Rosenhaus:2014zza}, see also \cite{Casini:2014yca} for interpretation based on calculations of the mutual information in Minkowski space. Our findings here point out in favour of the difference between the minimally and conformally coupled scalar fields in the limit of flat space.

In section \ref{sec:spec} we derive the spectral representation of entanglement entropy flow. The final expression \reef{Sprime} depends on a specific spectral function which determines the spin-0 part of a two-point function of the energy-momentum tensor  \cite{Osborn:1999az}. Reflection positivity ensures that the spectral function is positive, and therefore \reef{Sprime} suggests that the rate of change of entanglement entropy along the RG trajectory is negative provided that the integral on the right hand side converges. In particular, it implies $S_\mt{EE}^\mt{UV}\geq S_\mt{EE}^\mt{IR}$ for a finite part of EE in $d=3$. 

Obviously, this inequality is not the same as \reef{Fthm}, and therefore it does not prove the $F$-theorem \cite{Jafferis:2011zi,Klebanov:2011gs} in three dimensions. To understand the difference between the two inequalities, it is enough to consider the universal `area law'  for a massive free field, see Appendix \ref{freeflat}. In $d=3$ it represents a finite contribution to EE which grows linearly in the IR limit. Since massive degrees of freedom decouple in the deep IR, the universal `area law' becomes part of the UV physics, and therefore it should be subtracted to isolate a `true' universal entanglement entropy in this limit. 

Of course, IR entanglement entropy may contain all kind of finite terms associated with the fingerprints of UV physics. Such remnants of the RG trajectory should be subtracted to extract the universal piece. REE is a particular subtraction scheme which proved to be effective in building a c-function in Minkowski space \cite{Casini:2012ei}. However, we have shown that it is not as good on a sphere. Unfortunately, we were not able to identify a reasonable candidate for a $c$-function on a sphere. Thermodynamic inequalities might be a good source to look for plausible candidates, \eg the generalized second law is one such example \cite{Wall:2009wm}. We will explore this avenue elsewhere.

\acknowledgments  
We thank Igor Klebanov, Robert C. Myers and Vladimir Rosenhaus for helpful discussions. The work of MS is supported in part by NSF Grant PHY-1214644 and the Berkeley Center for Theoretical Physics. The work of OB and DC is partially supported by the Israel Science Foundation (grant 1989/14 ), the  US-Israel bi-national fund (BSF) grant 2012383 and the German Israel bi-national fund GIF grant number I-244-303.7-2013.

\appendix

\section{Universal area law}
\labell{freeflat}

In has been shown in \cite{Rosenhaus:2014ula,Casini:2014yca} that variation of the entanglement entropy with respect to the relevant coupling can be expressed in terms of a particular structure in the spectral decomposition of the energy-momentum tensor
\be
 \lambda{\del S_\mt{EE} \over \del\lambda}={-4\pi \, \Omega_{d-1} \A_\Sigma\over 2^{d}(d-\Delta-\beta_\lambda)(d-1)(d+1)\Gamma(d)} \int_0^\infty d\mu \, c^{(0)}(\mu)~,
 \labell{SprimeFlat}
\ee
where $c^{(0)}(\mu)$ is the spectral function which corresponds to spin $s=0$ states in Minkowski space \cite{Cappelli:1990yc}.  For free conformally coupled scalar and Dirac fermion, we have \cite{Cappelli:1990yc}
\bea
  c^{(0)}_F(\mu)&=&2^{[d/2]}\, {2(d+1)(d-1)\over \Omega_{d-1}^2} \, m^2 \, \mu^{d-5}\(1-{4m^2\over \mu^2}\)^{(d-1)/2}\Theta(\mu-2m)~,
  \non
  c^{(0)}_S(\mu)&=&{8(d+1)(d-1)\over \Omega_{d-1}^2}\,m^4 \, \mu^{d-7}\(1-{4m^2\over \mu^2}\)^{(d-3)/2}\Theta(\mu-2m)~.
  \labell{spec0}
\eea
Substituting into \reef{SprimeFlat}, and using dimensional regularization, yields
\bea
m{\del S_\mt{EE}^{\mt{Dirac}} \over \del m} &=&-{2^{\[{d\over 2}\]}\over 12} {\Gamma\( {4- d\over 2}\) \over (4\pi)^{d-2\over 2}} A_{\Sigma}\, m^{d-2}
 \\[1em]
 m^2{\del S_\mt{EE}^{\mt{scalar}} \over \del m^2} &=&  {(d-4)\over 24(d-1)} {\Gamma\( {4- d\over 2}\) \over (4\pi)^{d-2\over 2}} A_{\Sigma}\, m^{d-2}%=
\eea
Hence, we get 
\be
m {\del S_\mt{univ}^{\mt{Dirac}}\over \del m}= \left\lbrace \begin{matrix}
   {   (-)^{{d\over 2}} (d-2)  \over  6(2\pi)^{d-2\over 2} \Gamma\({d\over 2}\)} A_{\Sigma}\ m^{d-2}\,\log(m\delta) &
      \qquad {\rm for\ even}\ d\,,   \\[1em]
    {   (-)^{d-1\over 2}\pi (d-2)  \over 12 \sqrt{2}(2\pi)^{d-2\over 2} \Gamma\({d\over 2}\)}  A_{\Sigma}\ m^{d-2}\hfill  &
      \qquad {\rm for\ odd}\ d\,,\hfill
\end{matrix} \right.
\ee
where simple pole in the gamma function is identified with $\log(m\delta)$ \footnote{To establish this dictionary, it is enough to introduce a sharp cut off $\mu_\mt{max}=1/\delta$ in \reef{SprimeFlat}.}.

Similarly for the conformally coupled scalar,
\be
m^2 {\del S_\mt{univ}^{\mt{scalar}}\over \del m^2}= \left\lbrace \begin{matrix}
   {(d-2)(d-4)\over 24(d-1)} {   (-)^{{d\over 2}+1}  \over  (4\pi)^{d-2\over 2} \Gamma\({d\over 2}\)} A_{\Sigma}\ m^{d-2}\,\log(m\delta) &
      \qquad {\rm for\ even}\ d\,,   \\[1em]
    {(d-2)(d-4)\over 48(d-1)} {   (-)^{d+1\over 2}\pi  \over  (4\pi)^{d-2\over 2} \Gamma\({d\over 2}\)}  A_{\Sigma}\ m^{d-2}\hfill  &
      \qquad {\rm for\ odd}\ d\,.\hfill
\end{matrix} \right.
\ee
Note that this result is different from the universal `area law' for minimally coupled scalar field \cite{Hertzberg:2010uv,Huerta:2011qi,Lewkowycz:2012qr},
\be
m^2 {\del S_\mt{univ}^{\mt{scalar}}\over \del m^2}= \left\lbrace \begin{matrix}
	{(d-2)\over 12} {   (-)^{{d\over 2}}  \over  (4\pi)^{d-2\over 2} \Gamma\({d\over 2}\)} A_{\Sigma}\ m^{d-2}\,\log(m\delta) &
	\qquad {\rm for\ even}\ d\,,   \\[1em]
	{(d-2)\over 24} {   (-)^{d-1\over 2}\pi  \over  (4\pi)^{d-2\over 2} \Gamma\({d\over 2}\)}  A_{\Sigma}\ m^{d-2}\hfill  &
	\qquad {\rm for\ odd}\ d\,.\hfill
\end{matrix} \right.
\ee

\section{Free fields on a deformed waveguide}
\labell{curvEE}

In this Appendix we perform an independent computation of the leading curvature correction displayed in \reef{Sphi} and \reef{Spsi}. To this end one can resort to the replica trick \cite{Callan:1994py,Fursaev:1995ef,Calabrese:2004eu,Calabrese:2005zw} combined with the heat kernel technique, see \eg \cite{Vassilevich:2003xt}. Indeed, rotational symmetry around the entangling surface makes it possible to introduce a well-defined conical defect \cite{Dowker:2010yj,Dowker:2012rp,Dowker:2014zwa}. 
However, we choose to follow a different approach \cite{Rosenhaus:2014woa}, where a special role played by the neighbourhood of the entangling surface is made manifest. 

Recall that divergences of the entanglement entropy are local and dominated by the region near the entangling surface. In particular, to recover \reef{Sphi} and \reef{Spsi} it is enough to explore our metric \reef{foliation} in the vicinity of the entangling surface at $\theta=\pi/2$ 
\be
 ds^2=(1-{r^2\over 3\,R^2}+\ldots)r^2\,d\tau^2 + R^2 d\theta^2+(1-{r^2\over R^2}+\ldots)R^2\,d\Omega_{d-2}^2  ~,
\ee
where $r\equiv R(\theta-\pi/2)$. To leading order in the radial distance $r$, the geometry can be approximated by a waveguide with spherical cross-section $R^2\times S^{d-2}$, whereas subleading terms can be treated as perturbations.

In Cartesian coordinates $x^1=r\cos\theta ~, x^2=r\sin\theta$, the expansion of the metric takes the form
\be
 ds^2=(1-{1\over 3}\cR_{abcd}x^b\,x^d+\ldots)dx^a dx^c +(\gamma_{ij}+\cR_{iacj}x^ax^c+\ldots)dy^idy^j  ~,
 \labell{metric}
\ee
where ellipsis encode higher derivative terms which are irrelevant for our needs, $\{y^i\}_{i=1}^{d-2}$ and $\gamma_{ij}$ are coordinates and induced metric on $S^{d-2}$ respectively, and $ \cR_{\mu\nu\sigma\rho}$ is the Riemann curvature tensor on $S^d$
\be
 \cR_{\mu\nu\rho\sigma}={1\over R^2} (g_{\mu\rho}g_{\nu\sigma}-g_{\mu\sigma}g_{\nu\rho})~.
\ee
Of course, for a generic entangling surface \reef{metric} contains terms with extrinsic curvatures \cite{Lewkowycz:2013nqa,Fur13,Bhattacharyya:2013gra,Dong:2013qoa,Camps:2013zua,Rosenhaus:2014woa}, \footnote{See also critique of \cite{Dong:2013qoa,Camps:2013zua} in \cite{Bhattacharyya:2014yga}.} however in our setup they vanish due to rotational symmetry around $\Sigma$. 

Next we observe that the curvature terms in \reef{metric} are small close to the entangling surface ($x^a=0$) where the divergences are localized, and therefore we treat them as perturbations, $h_{\mu\nu}$, of the waveguide geometry, see \cite{Rosenhaus:2014woa} for details. In particular, it follows from \reef{flow} that linear correction to the entanglement entropy takes the form 
\be
 \delta S_\mt{EE}= {1\over 2} \int d^{2\,} x \int d^{d-2}y\sqrt{\gamma} \, \langle T^{\mu \nu}(x,y) K\rangle  h_{\mu\nu}(x,y) + \mathcal{O}(h^2) ~.
\ee
Only universal divergences of the above expression can be taken at face value, otherwise there is no reason to expect that higher order terms in $h_{\mu\nu}$ do not contribute. 

Now the first term within parenthesis in \reef{Sphi}, \reef{Spsi} represents the universal `area law' \cite{Hertzberg:2010uv,Huerta:2011qi,Lewkowycz:2012qr}, see also \cite{Rosenhaus:2014ula,Casini:2014yca} and Appendix \ref{freeflat}. This term is entirely encoded in the leading order entanglement entropy, whereas $\delta S_\mt{EE}$ contributes to the second term in \reef{Sphi}, \reef{Spsi} which is proportional to the Riemann tensor. Given that $h_{\mu\nu}$ is also proportional to the Riemann tensor, we can replace $\langle T^{\mu \nu} K\rangle$ by its flat space counterpart without losing contributions to \reef{Sphi}, \reef{Spsi}, then according to \cite{Rosenhaus:2014ula}
\bea
\delta S_\mt{EE} &=&{\pi \, \Omega_{d-1} \over 2^{d-2}(d-1)^2 (d+1) \Gamma(d)} \int_{\Sigma} \( \delta^{ac}\delta^{ij}\cR_{iajc} +{1\over 2}\delta^{ac}\delta^{bd}\cR_{abcd} \)
\int_0^\infty {d\mu \over \mu^2} c^{(0)}(\mu)
 \label{varEE2}
 \\[1em]
 &+&{\pi \, \Omega_{d-1} \over 2^{d-2}(d-1)(d+1) \Gamma(d)} \int_{\Sigma} \({d-2\over 2}\delta^{ac}\delta^{bd}\cR_{abcd} - \delta^{ac}\delta^{ij}\cR_{iajc}\)\int_0^\infty {d\mu\over \mu^2} c^{(2)}(\mu)~,
\nonumber
\eea
where $c^{(0)}(\mu)$ and $c^{(2)}(\mu)$ are two spectral functions which define a two point function of the energy-momentum tensor for a generic QFT in Minkowski space \cite{Cappelli:1990yc}. For free fields $c^{(0)}(\mu)$ is given by \reef{spec0}, whereas $c^{(2)}(\mu)$ reads \cite{Cappelli:1990yc}
\bea
 c^{(2)}_F(\mu)&=&2^{[d/2]}\, {(d-1)\over 2 \, \Omega_{d-1}^2} \, \mu^{d-3}\(1-{4m^2\over \mu^2}\)^{(d-1)/2}\(1+{2\over d-1}{4m^2\over \mu^2}\)\Theta(\mu-2m)~,
 \non[1em]
 c^{(2)}_S(\mu)&=& {\mu^{d-3}\over \Omega_{d-1}^2} \(1-{4m^2\over \mu^2}\)^{(d+1)/2}\Theta(\mu-2m)~.
\eea

Hence,
\bea
 \delta S_\mt{univ}^\mt{scalar}&=&
 \frac{(-)^{\frac{d-2}{2}}(d-2)(-76+84d-25d^2+2d^3)}{240(4\pi)^{\frac{d-2}{2}}(d-1)^2 \Gamma(\frac{d}{2})} \, \frac{A_\Sigma m^{d-4}}{R^2} \, \log (m \delta) ~,
 \non[1em]
 \delta S_\mt{univ}^\mt{Dirac} &=& 
  \frac{(-)^{\frac{d}{2}}(d-2)(d-3)}{60(2\pi)^{\frac{d-2}{2}} \Gamma(\frac{d}{2})} \, \frac{A_\Sigma m^{d-4}}{R^2} \, \log (m \delta)~,
\eea
where we introduced a UV cut off $\mu_\mt{max}\sim 1/\delta$. This contribution should be combined with the leading order entanglement entropy. To leading order the geometry is identical to a waveguide with spherical cross-section, and therefore for Dirac fermion we can use the results of \cite{Lewkowycz:2012qr}
\be
 S_\mt{univ}^\mt{Dirac}\Big|_{R^2\times S^{d-2}}
 =\frac{ (-1)^{\frac{d}{2}} A_\Sigma }{6(2\pi)^{\frac{d-2}{2}}\Gamma(\frac{d}{2})} \( m^{d-2} + \frac{(d-2)^2(d-3)}{24}\, {m^{d-4} \over R^2} + \ldots \)  \log (m \delta)~,
\ee 
while for conformally coupled scalar, the result reads\footnote{Entanglement entropy of a scalar field on a waveguide geometry was studied in \cite{Lewkowycz:2012qr} using the heat kernel methods. In particular, the authors apply the same heat kernel, $K_{\mathcal{C}_n}$, on a two-dimensional cone,  $\mathcal{C}_n$, with an angular excess $2\pi(n-1)$ to evaluate entanglement entropy for both minimally and non-minimally coupled scalar fields. However, to get \reef{swave} we make use of the following expression
\be
 \text{Tr} K_{\mathcal{C}_n}(t) = {(1-n)\over 6} (1-6\xi)+n \text{Tr} K_{R^2}(t) + \mathcal{O} (1-n)^2 ~,
\ee 
where $\xi$ represents a non-minimal coupling to the background curvature. To derive it, there is no need to find or use the exact solution \cite{Dowker:1977zj,Dowker:1987pk,Deser:1988qn} to the heat kernel equation on a cone. Instead, one can use the general form of the heat kernel expansion in the vicinity of $t=0$ \cite{Vassilevich:2003xt}, and substitute curvature scalar on $\mathcal{C}_n$
\be
 \cR=4\pi(1-n)\delta_\Sigma + \mathcal{O} (1-n)^2 ~,
\ee
where $\delta_\Sigma$ is a two-dimensional $\delta$-function supported on $\Sigma$. 
}
\be
 S_\mt{univ}^\mt{scalar}\Big|_{R^2\times S^{d-2}}
 =\frac{ (-1)^{\frac{d}{2}} (1-6\xi_c)A_\Sigma }{6(4\pi)^{\frac{d-2}{2}}\Gamma(\frac{d}{2})} \( m^{d-2} - \frac{(d-2)^2(d-3)(1-6\xi_c)}{12}\, {m^{d-4} \over R^2} 
 + \ldots \)  \log (m \delta)~,
 \labell{swave}
\ee
 
Combining altogether, we obtain 
\bea
 S_\mt{univ}^\mt{scalar}&=&S_\mt{univ}^\mt{scalar}\Big|_{R^2\times S^{d-2}}+ \delta S_\mt{univ}^\mt{scalar}
 ={(d_0-4)\over 12(d-1)} { (-)^{d_0/2+1}  A_{\Sigma} \over  (4\pi)^{(d_0-2)/2} \Gamma\({d_0\over 2}\)} 
 \non
 &&\quad\quad\quad\quad\quad\quad\quad
  \times\(  m^{d_0-2} + \frac{(d_0-2)(d_0-6) \left(5 d_0^2-18 d_0+4\right)}{120(d-4)}\frac{m^{d_0-4}}{R^{2}}\)\log (m \delta)~,
\non[1em]
 S_\mt{univ}^\mt{Dirac}&=&S_\mt{univ}^\mt{Dirac}\Big|_{R^2\times S^{d-2}}+ \delta S_\mt{univ}^\mt{Dirac}
 \\
 &=&\frac{ (-1)^{\frac{d}{2}} A_\Sigma }{6(2\pi)^{\frac{d-2}{2}}\Gamma(\frac{d}{2})} \( m^{d-2} + \frac{(d-2)(d-3)(5d+2)}{120}\, {m^{d-4} \over R^2} + \ldots \)  \log (m \delta)~,
 \nonumber
\eea
in full agreement with \reef{Sphi}, \reef{Spsi}.

\section{Addition theorem for spherical harmonics on $S^3$}
\label{appx:add-sph-harmon}

In this appendix we review the proof of the following useful mathematical result
 \be
 Y_{l_300}(\gamma)={\sqrt{2}\pi\over l_3+1}\sum_{l_2=0}^{l_3}\sum_{l_1=-l_2}^{l_2}
 Y_{l_3l_2l_1}^{*}(y)Y_{l_3l_2l_1}(x)
 \quad ,
 \label{eqn:addition-theorem}
 \ee
where $x,y$ are two arbitrary points on $S^{\,3}$ and $\gamma$ is an
angle between them. This identity is called the addition theorem for
spherical harmonics.

Any function $f(x)$ on $S^{\,3}$ can be expanded as follows
 \be
 f(x)=\sum_{l_3=0}^{\infty}\sum_{l_2=0}^{l_3}\sum_{l_1=-l_2}^{l_2} A_{l_3l_2l_1} Y_{l_3l_2l_1}(x)
 \quad, \quad A_{l_3l_2l_1}=\int_x Y_{l_3l_2l_1}^{*}(x)f(x)
 \quad .
 \ee
If we set $\chi=0$ then according to
(\ref{eqn:spherical-harmonics}),(\ref{eqn:associated-Legandre}) only
terms with $l_2=0\Rightarrow l_1=0$ contribute
 \be
 f(x)|_{\chi=0}=\sum_{l_3=0}^{\infty}A_{l_300}{l_3+1 \over \sqrt{2}\pi}
 \quad, \quad A_{l_300}=\int_x Y_{l_300}(x)f(x)
 \quad .
 \label{eqn:A3}
 \ee

On the other hand, expanding $Y_{l_300}(\gamma)$ in spherical
harmonics and taking into account that it is a spherical harmonic of
order $l_3$, yields
 \be
 Y_{l_300}(\gamma)=\sum_{l_2=0}^{l_3}\sum_{l_1=-l_2}^{l_2} A_{l_3l_2l_1}(y) Y_{l_3l_2l_1}(x)
 \quad, \quad A_{l_3l_2l_1}(y)=\int_x Y_{l_3l_2l_1}^{*}(x)Y_{l_300}(\gamma)
 \quad .
 \ee
$A_{l_3l_2l_1}(y)$ can be viewed as $A_{l_300}$ coefficient in an
expansion of $Y_{l_3l_2l_1}^{*}(x)$ in a series of
$Y_{l_3l_2l_1}(\gamma)$ referred to the axis $y$. From
(\ref{eqn:A3}) it is then found that, since only one $l_3$ is
present this coefficient is given by
 \be
 A_{l_3l_2l_1}(y)={\sqrt{2}\pi\over l_3+1}Y_{l_3l_2l_1}^{*}(y)
 \quad ,
 \ee
from which (\ref{eqn:addition-theorem}) follows.

\bibliographystyle{utphys}

\bibliography{lib}

\end{document}